\documentclass[%
 twocolumn,
 amsfonts,amsmath,amssymb,
 aps,
prd,
floatfix,
showpacs,
longbibliography
]{revtex4-2}

\usepackage{graphicx}
\usepackage{dcolumn}
\usepackage{bm}
\usepackage{wasysym}
\usepackage[dvipsnames]{xcolor}
\definecolor{ReflexBlue}{rgb}{ .0902,.0902,.5882}
\usepackage{hyperref}
\hypersetup{
    colorlinks=true,
    linkcolor=ReflexBlue,
    filecolor=magenta,      
    urlcolor=ReflexBlue,
    citecolor=ReflexBlue,
}
\let\Phi\varPhi

\begin{document}


\title{Geometry near the inner horizon of a rotating, accreting black hole}

\author{Tyler McMaken}
 \email{Tyler.McMaken@colorado.edu}
\author{Andrew J. S. Hamilton}
 \email{Andrew.Hamilton@colorado.edu}
\affiliation{%
 JILA and Dept.\ Physics, U.\ Colorado Boulder, CO 80309, USA
}%
\date{\today}

\begin{abstract}
Here we present a novel classical model to describe the near-inner horizon geometry of a rotating, accreting black hole. The model assumes spacetime is homogeneous and is sourced by radial streams of a collisionless, null fluid, and it predicts that the standard Poisson-Israel mass inflation phenomenon will be interrupted by a Kasner-like collapse toward a spacelike singularity. Such a model is shown to be valid at the inner horizon of astrophysically realistic black holes through comparison to the conformally-separable model, which provides a natural connection of the Kerr metric to a self-similar, accreting spacetime. We then analyze the behavior of null geodesics in our model, connecting them to the Kerr metric in order to answer the practical question of what an infalling observer approaching the inner horizon might see.
\end{abstract}

\maketitle


\section{\label{sec:introduction}Introduction}

Physicists have long wondered what happens inside astrophysically realistic black holes. The exterior geometry of a back hole has been well-established to be described completely by the Kerr-Newman metric, since any perturbations during a generic collapse will be radiated away and leave the black hole with only three uniquely-identifying parameters: a charge $Q$, angular momentum $J$, and mass $M$. But when the Kerr-Newman solution is extended to the interior of a black hole's event horizon, some puzzling, nonphysical structures emerge.

Within the simplest model of black holes (the Schwarzschild solution), no major peculiarities or nonphysical structures arise except the divergence of the spacetime curvature at $r$=0 at a spacelike singularity. Prior to the 1960s, many opposed the idea that a realistic gravitational collapse would lead to a singularity, since most known models of collapse were highly idealized and unstable to perturbations \cite{lif63}. However, in 1965, Penrose published a theorem demonstrating the inevitability of singularities within event horizons of black holes \cite{pen65}, and soon after, Belinskii, Khalatnikov, and Lifschitz found a realistic model for such a collapse to a spacelike singularity (the so-called BKL collapse), which is highly complex and oscillatory \cite{bel70}.

In spite of these efforts, the fact remains that most, if not all, black holes are not spherically symmetric and instead carry at least some angular momentum. The structure of the Kerr-Newman interior differs drastically from that of a Schwarzschild black hole---instead of a spacelike singularity at the center, the Kerr-Newman solution has a timelike singularity along with a second horizon within the event horizon. This inner horizon coincides with the singularity when $J=0$ and $Q=0$, but for nonzero spin or charge, the spacetime between the inner horizon and the singularity forms a region in which predictability breaks down---general relativity is powerless to predict unambiguously what would happen if an observer passes through the inner horizon, because such an observer would be able to view the singularity.

Aside from the breakdown of predictability, the added interior structure of a rotating or charged black hole is problematic for another reason. As first pointed out by Penrose in the context of a Reissner-Nordstr{\"o}m (charged) black hole, the inner horizon is a surface of infinite blueshift, so that an infalling observer at an inner horizon would see the entire history or future of the Universe flash before their eyes as the energy of any incoming radiation becomes classically unbounded \cite{pen68}. Penrose conjectured that the added effects of this diverging radiation would change the underlying spacetime curvature of the vacuum solution \cite{sim73}. This conjecture was finally confirmed a few decades later, when Poisson and Israel performed a full nonlinear perturbation analysis in a seminal 1990 paper \cite{poi90,bar90}. Poisson and Israel concluded that the crossing of ingoing and outgoing shells of null dust at the inner horizon would lead to a divergence of the spacetime curvature. Poisson and Israel dubbed this divergence ``mass inflation,'' because an observer near the inner horizon would measure an exponentially large quasi-local internal mass parameter (though the global mass as measured at infinity would remain finite). The observer near the inner horizon would then see an asymptotically Schwarzschild-like geometry with an enormous Schwarzschild mass $M$, and the journey to the inner horizon would encompass all but the last Planck time of the black hole's classical history.

Classically, the Poisson-Israel toy model of mass inflation leads to the formation of a null weak curvature singularity, in which the curvature locally diverges but the tidal distortion of extended objects travelling along timelike geodesics remains finite, allowing for the continuation of spacetime beyond the Cauchy surface \cite{ori91,ori92}. Dafermos extended this result for the less-simplified Einstein-Maxwell-real scalar field equations \cite{daf05}, and Ori and others found that null curvature singularities provide a generic class of solutions to the Einstein equations, adding it to the list of known possible singularities that had previously only included the BKL singularity \cite{bra95,ori96,ori98}. The BKL and null curvature singularities are quite different in nature---though they both may be oscillatory in nature, the BKL singularity is strong and spacelike, whereas the null curvature singularity is weak and lightlike \cite{ori99}.

Despite the enticement of the null weak curvature singularity, both in its mathematical accessibility and in its potential to allow for a gateway to another Universe, it suffers one fatal flaw. One of the key assumptions for all the models that predict a null weak singularity is that the collapsed black hole is in complete isolation. Under this assumption, which still dominates much of the research program for mass inflationary phenomena to this day \cite{rub21}, the only source of ingoing perturbations is the Price tail, a stream of gravitational waves emitted and backscattered during the collapse. The Price tail decays with an inverse power law in advanced time, and calculations for the formation of a null weak singularity assume that no additional radiation perturbs the metric above that power law \cite{pri72}. However, astrophysical black holes continue to accrete material long after the initial gravitational collapse, and even the cosmic microwave background radiation would dominate over the longest-lived Price tail modes of a stellar-mass black hole after only 1 second \cite{ham17}.

Motivated by this shortcoming, Burko found numerically that a null weak singularity only forms for a sufficiently steep radiation power law drop-off, and that if it does not drop off quickly enough, a spacelike singularity will form at the intersection of the ingoing and outgoing inner horizons and grow until it has completely sealed off the Kerr tunnel \cite{bur02,bur03}.

\begin{figure}
    \includegraphics[width=\columnwidth]{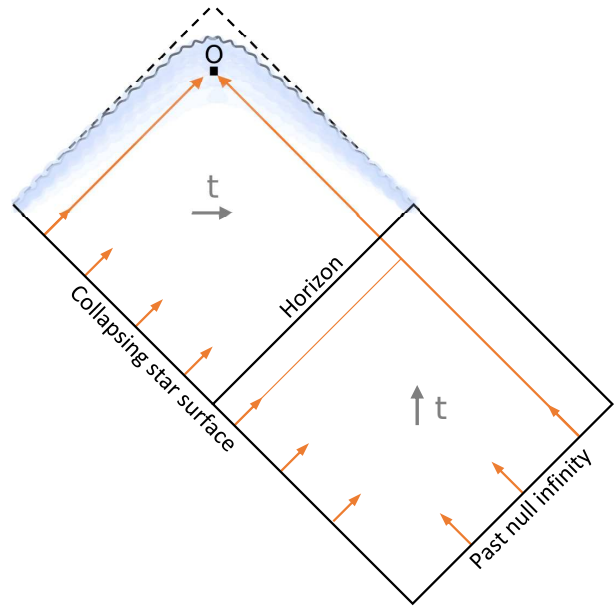}
    \caption{\label{fig:penrose} (Color online). Penrose diagram for the late-time evolution of a collapsed star with a Kerr exterior (white) matched to an inflationary Kasner regime (shaded blue). The inner horizons (dashed lines) of the Kerr metric are superseded by the BKL singularity of the inflationary Kasner model (squiggled line). The gray arrows labelled $t$ indicate the direction of increasing Boyer-Lindquist time.}
\end{figure}

Hamilton subsequently developed a self-similar model for the inner horizon spacetime that generalizes the mass inflation phenomenon to include arbitrary ingoing and outgoing collisionless streams of radiation at arbitrary times, first for spherical-symmetric spacetimes \cite{ham10} and soon after for the more realistic case of rotating black holes \cite{ham11a,ham11b,ham11c}. The rotating case, which assumes conformal separability, is reviewed in more detail in Section \ref{subsec:cskerr}. The key conclusion from this model is that the continued streams eventually slow the inflation of the curvature, causing the spacetime to collapse radially. The resulting global geometry, used throughout this paper, is shown in the Penrose diagram of Fig.~\ref{fig:penrose}.

Though the conformally-separable model of Hamilton is valid through the inflation and the beginning of the subsequent collapse of the spacetime, it eventually fails once the rotational motion of the streams becomes comparable to their radial motion. After this point, numerical calculations indicate that the collapse follows a BKL-like behavior \cite{ham17}. However, recent semiclassical calculations suggest that quantum backreaction effects may alter or even invert the collapse \cite{bar20}. Thus, one may wish to calculate the renormalized stress-energy tensor in the conformally-separable Kerr spacetime. Such a calculation has not yet been attempted because of the complexity of Hamilton's model; however, here we derive a new model that considerably simplifies the conformally-separable Kerr model while still retaining its essential features of inflation and collapse. This model, which we call the inflationary Kasner model, will be shown to provide a reasonable continuation of the Kerr metric near the inner horizon through the first two Kasner epochs of its BKL collapse, and it will hopefully allow for future quantum calculations in this region.

In Section~\ref{sec:ikasner}, the inflationary Kasner metric is derived as a general solution to Einstein's equations for a homogeneous spacetime sourced by a null, perfect fluid. In particular, we find that for a line element of the form
\begin{equation}
    ds^2=-\alpha(T)^2dT^2+a_1(T)^2dx^2+a_2(T)^2dy^2+a_3(T)^2dz^2,
\end{equation}
the general solution for an energy-momentum tensor whose only non-negligible components are ${T_{00}=T_{11}}$ takes the form
\begin{align}
    a_1&\propto T^{-p/2}\exp\left(T^{2p}\right),\nonumber\\
    a_2&\propto T^p,\nonumber\\
    a_3&\propto T^p,
\end{align}
for some arbitrary constant $p$. This metric is dubbed the inflationary Kasner metric because it is a natural non-vacuum extension of the Kasner metric, a vacuum solution used to model collapse to a BKL singularity. During a BKL collapse, the spacetime undergoes a series of ``BKL bounces,'' between which the evolution is described by the Kasner metric's power law behavior. The BKL model is described in more detail in Sec.~\ref{subsec:interpretation}, in which we show how the two epochs of the inflationary Kasner solution can be reduced in certain limits to previously obtained results.

Then, in Sec.~\ref{sec:application}, we show how the inflationary Kasner model can be applied to the inner geometry of astrophysical black holes. To do so, we connect our model to the Kerr metric in a regime where both are valid, employing Hamilton's conformally-separable Kerr model to facilitate the matching and to determine the degree to which the assumptions of the inflationary Kasner model are valid near the inner horizon. Sec.~\ref{subsec:cskerr} is devoted to reviewing the conformally-separable Kerr model and comparing it to the inflationary Kasner model, where we find that the inflationary exponent $\xi$ of the conformally-separable model is related to the inflationary Kasner time $T$ via the relation ${T^{p}\propto\text{ e}^{-\xi}}$. Then, in Sec.~\ref{subsec:geodesics}, we analyze the behavior of null geodesics in our model, ray-tracing them from an observer in the inflationary Kasner spacetime backwards until they connect to null geodesics in the Kerr spacetime. Such a matching allows us in Sec.~\ref{sec:obs_experience} to answer the practical question of what an observer falling toward the inner horizon of an astrophysical, classical black hole might see.

\section{\label{sec:ikasner}Inflationary Kasner metric}

\subsection{\label{subsec:preliminaries}Preliminaries}

Throughout this paper, we use the metric signature ${-+++}$ and geometric units where ${c=G=1}$.

In our analysis we use an orthonormal tetrad formalism, in which quantities are defined in the tetrad basis ${\{e_{\hat{m}}\}}$ to yield physically measured components in the local, Cartesian frame of an observer. In such a formalism, coordinate-frame quantities can be converted into tetrad-frame quantities through the vierbein ${e^{\hat{m}}_{\hspace{5pt}\mu}}$, which can be read off directly from a line element via
\begin{equation}
    ds^2=g_{\mu\nu}dx^{\mu}dx^{\nu}=\eta_{\hat{m}\hat{n}}e^{\hat{m}}_{\hspace{5pt}\mu}e^{\hat{n}}_{\ \nu}dx^\mu dx^\nu.\label{eq:def_vierbein}
\end{equation}

Indices for abstract, Einstein-summed tetrad-frame quantities are denoted by lowercase Latin letters with hats, while indices for abstract, Einstein-summed coordinate-frame quantities are denoted by lowercase Greek letters. Indices for specific components of tetrad-frame quantities are denoted by Arabic numerals, while those of coordinate-frame quantities are given by their standard Latin or Greek letters. Thus a tetrad-frame four-vector can be expressed as ${k^{\hat{m}}=\{k^0,k^1,k^2,k^3\}}$, a coordinate-frame one as ${k^\mu=\{k^t,k^r,k^{\theta},k^{\phi}\}}$, and the conversion between the two is given by
\begin{equation}
    k^{\hat{m}}=e^{\hat{m}}_{\hspace{5pt}\mu}k^\mu.
\end{equation}

For a more complete review of the tetrad formalism, see M{\"u}ller's ``Catalogue of Spacetimes'' \cite{mul09} or Chandrasekhar's \textit{The Mathematical Theory of Black Holes} \cite{cha83}.

\subsection{\label{subsec:ikasner}Derivation of the line element}

The purpose of this subsection is to derive the line element for the inflationary Kasner metric, which in its final form reads
\begin{equation}
    ds^2=a_1^2\left(-dT^2+dx^2\right)+a_2^2\left(dy^2+dz^2\right),\label{eq:ikasner}
\end{equation}
with
\begin{align}\label{eq:ikasner_scalefactors}
    a_1&=\frac{\text{e}^{(T-T_0)/2}}{\sqrt{16\pi\Phi_0T_0}}\left(\frac{T}{T_0}\right)^{-1/4},\nonumber\\ a_2&=\frac{1}{\sqrt{16\pi\Phi_0T_0}}\left(\frac{T}{T_0}\right)^{1/2},
\end{align}
where the time coordinate $T$ progresses backward from the positive constant $T_0$ to $0$, and the positive constant $\Phi_0$ represents the mass-energy density of the streams of fluid seen by an observer at ${T=T_0}$. In general, the mass-energy density will be found to depend on $T$ through the relation
\begin{equation}
    \Phi(T)=\frac{\Phi_0}{\sqrt{T/T_0}\text{ e}^{T-T_0}}.\label{eq:ikasner_Phi}
\end{equation}

The form of the inflationary Kasner line element of Eq.~(\ref{eq:ikasner}) relies on two main assumptions. First, assume the metric is spatially homogeneous, so that the metric coefficients are functions only of the time coordinate $T$. Such a requirement exists in a more stringent form for the Kasner metric described in the next section, in which the metric coefficients are power laws in $T$ during a BKL collapse. Second, assume the solution to Einstein's equations is sourced by a collisionless, null, perfect fluid in the radial direction. In a tetrad frame, such a source corresponds to the energy-momentum tensor
\begin{equation}\label{eq:ikasner_Tmn}
     T_{\hat{m}\hat{n}}=
        \begin{pmatrix}
        \Phi & 0 & 0 & 0\\
        0 & \Phi & 0 & 0\\
        0 & 0 & 0 & 0\\
        0 & 0 & 0 & 0
        \end{pmatrix},
\end{equation}
where $\Phi$ is the mass-energy density of the null streams. For a realistic accreting black hole, even if the accreted material is not null and purely radial, near the inner horizon, all streams of matter and radiation are expected to focus along the principal null directions ultrarelativistically, so that the energy-momentum tensor to a good approximation takes the form above.

Assume the line element (and therefore the vierbein) can be written in a diagonal basis. Thus, the tetrad ${\text{1-forms}}$ may be written as
\begin{subequations}\label{eq:gen_vierbein}
\begin{align}
    e^0_{\ \mu}dx^\mu&=\frac{a_1a_2a_3}{T}dT,\\
    e^1_{\ \mu}dx^\mu&=a_1dx,\\
    e^2_{\ \mu}dx^\mu&=a_2dy,\\
    e^3_{\ \mu}dx^\mu&=a_3dz,
\end{align}
\end{subequations}
where the scale factors $a_i$ are functions only of the time coordinate $T$ but are otherwise left arbitrary. The choice of the present form of $e^0_{\ T}$ will help to simplify later calculations; in general, $e^0_{\ T}$ may be any function of $T$ after a suitable transformation of the $T$ coordinate.

Instead of working in a coordinate basis and using the metric components to find the Christoffel connection coefficients $\Gamma^\mu_{\ \nu\rho}$, here we work entirely in a tetrad basis without reference to the coordinate frame, so we must first find the analogous tetrad connection coefficients. In the tetrad basis, the connection 1-forms ${\omega^{\hat{m}}_{\hspace{5pt}\hat{n}}}$ (which are antisymmetric in their tetrad-frame indices) can be defined by the torsion-free condition
\begin{equation}
    de^{\hat{m}}+\omega^{\hat{m}}_{\hspace{5pt}\hat{n}}\wedge e^{\hat{n}}=0,
\end{equation}
or in component form,
\begin{equation}
    \omega^{\hat{m}}_{\hspace{5pt}\hat{n}\rho}=e^{\hat{m}}_{\hspace{5pt}\mu}\nabla_\rho\hspace{2pt}e_{\hat{n}}^{\ \mu}.
\end{equation}

Converting all indices of the connection 1-form components to a tetrad basis then yields the Ricci rotation coefficients $\omega_{\hat{m}\hat{n}\hat{r}}$, antisymmetric in their first two indices:
\begin{equation}
    \omega_{\hat{m}\hat{n}\hat{r}}=\eta_{\hat{\ell}\hat{m}}e^{\ \rho}_{\hat{r}}\omega_{\ \hat{n}\rho}^{\hat{\ell}}=e_{\hat{m}}^{\hspace{5pt}\mu}e^{\ \rho}_{\hat{r}}\nabla_\rho\hspace{2pt}e_{\hat{n}\mu}.
\end{equation}
For the tetrad 1-forms of Eqs.~(\ref{eq:gen_vierbein}), the six nonvanishing Ricci rotation coefficients are as follows, where ${i\in\{1,2,3\}}$ and a dot above a variable indicates differentiation with respect to the time coordinate $T$:
\begin{equation}
    \omega_{0ii}=-\omega_{i0i}=\frac{T}{a_1a_2a_3}\frac{\dot{a}_i}{a_i}.
\end{equation}

The tetrad-frame Riemann curvature tensor components ${R_{\hat{k}\hat{\ell}\hat{m}\hat{n}}=e_{\hat{k}}^{\ \kappa}e_{\hat{\ell}}^{\ \lambda}e_{\hat{m}}^{\hspace{5pt}\mu}\left(\nabla_\kappa\nabla_\lambda-\nabla_\lambda\nabla_\kappa\right)e_{\hat{n}\mu}}$ can then be calculated, yielding 18 nonzero components: for ${i,j\in\{1,2,3\}}$ and ${i\ne j}$,
\begin{subequations}
\begin{align}
    R_{0i0i}&=R_{i0i0}=-R_{0ii0}=-R_{i00i}\nonumber\\
    &=\frac{T^2}{a_1^2a_2^2a_3^2}\left(\frac{\dot{a}_i}{a_i}\frac{d\ln\left(a_1a_2a_3/T\right)}{dT}-\frac{\ddot{a}_i}{a_i}\right),\\
    R_{ijij}&=-R_{ijji}=\frac{T^2}{a_1^2a_2^2a_3^2}\frac{\dot{a}_i\dot{a}_j}{a_ia_j}.
\end{align}
\end{subequations}

Then, the tetrad-frame Ricci tensor ${R_{\hat{k}\hat{m}}=\eta^{\hat{\ell}\hat{n}}R_{\hat{k}\hat{\ell}\hat{m}\hat{n}}}$, Ricci scalar ${R=\eta^{\hat{k}\hat{m}}R_{\hat{k}\hat{m}}}$, and tetrad-frame Einstein tensor ${G_{\hat{k}\hat{m}}=R_{\hat{k}\hat{m}}-\frac{1}{2}\eta_{\hat{k}\hat{m}}R}$ follow naturally. The resulting four nonzero Einstein components, where ${i\in\{1,2,3\}}$ with cyclic addition, are
\begin{subequations}\label{eq:ikasner_einstein}
\begin{align}
    G_{00}&=\frac{T^2}{a_1^2a_2^2a_3^2}\left(\frac{\dot{a}_1\dot{a}_2}{a_1a_3}+\frac{\dot{a}_1\dot{a}_3}{a_1a_3}+\frac{\dot{a}_2\dot{a}_3}{a_2a_3}\right),\\
    G_{ii}&=G_{00}-\frac{T}{a_1^2a_2^2a_3^2}\frac{d}{dT}\left(\left(\frac{\dot{a}_{i+1}}{a_{i+1}}+\frac{\dot{a}_{i+2}}{a_{i+2}}\right)T\right).
\end{align}
\end{subequations}

Under the assumption that the tetrad-frame energy-momentum tensor has the form of Eq.~(\ref{eq:ikasner_Tmn}), Einstein's equations give a system of four nontrivial equations:
\begin{subequations}\label{eq:ikasner_4syseq}
\begin{align}
    G_{00}&=8\pi\Phi,\label{eq:ikasner_4syseq_a}\\
    G_{11}&=8\pi\Phi,\label{eq:ikasner_4syseq_b}\\
    G_{22}&=0,\label{eq:ikasner_4syseq_c}\\
    G_{33}&=0.\label{eq:ikasner_4syseq_d}
\end{align}
\end{subequations}

The most natural solution to Eqs.~(\ref{eq:ikasner_einstein}) and (\ref{eq:ikasner_4syseq}) can be obtained by setting ${a_2=a_3}$, which reduces Eqs.~(\ref{eq:ikasner_4syseq_c}) and (\ref{eq:ikasner_4syseq_d}) to the same equation. Physically, this corresponds to the assumption that the $y$-$z$ plane, orthogonal to the streams of matter, remains isotropic, a reasonable assumption close to the horizon, given that any streams will focus ultrarelativistically in the $x$-direction. The remaining three equations simplify to
\begin{subequations}
\begin{align}
    8\pi\Phi&=\frac{T^2H_2}{a_1^2a_2^4}(2H_1+H_2),\label{eq:ikasner_3syseq_a}\\
    8\pi\Phi&=\frac{T^2H_2}{a_1^2a_2^4}\left(2H_1+H_2-\frac{2\dot{H}_2}{H_2}-\frac{2}{T}\right),\label{eq:ikasner_3syseq_b}\\
    0&=2H_1H_2+H_2^2-\frac{H_1+H_2}{T}-\dot{H}_1-\dot{H}_2,\label{eq:ikasner_3syseq_c}
\end{align}
\end{subequations}
where we have introduced the quantities
\begin{equation}
    H_i\equiv\frac{\dot{a}_i}{a_i}\qquad\implies\dot{H}_i=\frac{\ddot{a}_i}{a_i}-H_i^2
\end{equation}
for $i\in\{1,2\}$. Combining Eqs.~(\ref{eq:ikasner_3syseq_a}) and (\ref{eq:ikasner_3syseq_b}) to eliminate $\Phi$ yields
\begin{align}
   \frac{\dot{H}_2}{H_2}&=-\frac{1}{T}\nonumber\\
   \implies H_2&=\frac{p}{T}\nonumber\\
   \implies a_2&=C_2T^p,
\end{align}
where $p$ and $C_2$ are arbitrary integration constants. Substituting this solution into Eq.~(\ref{eq:ikasner_3syseq_c}) yields
\begin{align}
    \dot{H}_1&-\frac{(2p-1)}{T}H_1-\frac{p^2}{T^2}=0\nonumber\\
    \implies H_1&=-\frac{p}{2T}+qT^{2p-1}\nonumber\\
    \implies a_1&=C_1T^{-p/2}\exp\left(\frac{q}{2p}T^{2p}\right),
\end{align}
where $q$ and $C_1$ are arbitrary integration constants and the first-order differential equation in $H_1$ is most easily solved with the help of the integration factor ${\exp\left(-\int\frac{2p-1}{T}dT\right)=T^{1-2p}}$.

Thus, the tetrad 1-forms for the inflationary Kasner metric are
\begin{subequations}
\begin{align}
    e^0_{\ \mu}dx^\mu&=C_1C_2^2T^{3p/2-1}\exp\left(\frac{q}{2p}T^{2p}\right)dT,\\
    e^1_{\ \mu}dx^\mu&=C_1T^{-p/2}\exp\left(\frac{q}{2p}T^{2p}\right)dx,\\
    e^2_{\ \mu}dx^\mu&=C_2T^pdy,\\
    e^3_{\ \mu}dx^\mu&=C_2T^pdz.
\end{align}
\end{subequations}

Without loss of generality, replace the constants $C_1$, $C_2$, and $p$ through a set of redefinitions and coordinate transformations with the constants $T_0$ and $\Phi_0$, so that the vierbein becomes
\begin{subequations}\label{eq:ikasner_vierbein}
\begin{align}
    e^0_{\ \mu}dx^\mu&=\frac{\text{e}^{(T-T_0)/2}}{\sqrt{16\pi\Phi_0T_0}}\left(\frac{T}{T_0}\right)^{-1/4}\!dT,\label{eq:ikasner_vierbein_e0T}\\
    e^1_{\ \mu}dx^\mu&=\frac{\text{e}^{(T-T_0)/2}}{\sqrt{16\pi\Phi_0T_0}}\left(\frac{T}{T_0}\right)^{-1/4}\!dx,\\
    e^2_{\ \mu}dx^\mu&=\frac{1}{\sqrt{16\pi\Phi_0T_0}}\left(\frac{T}{T_0}\right)^{1/2}\!dy,\\
    e^3_{\ \mu}dx^\mu&=\frac{1}{\sqrt{16\pi\Phi_0T_0}}\left(\frac{T}{T_0}\right)^{1/2}\!dz.
\end{align}
\end{subequations}
These tetrad 1-forms lead directly to the line element of Eqs.~(\ref{eq:ikasner}) and (\ref{eq:ikasner_scalefactors}) through the relation in Eq.~(\ref{eq:def_vierbein}). This form is chosen to illuminate the physical meaning of the constants (as described in the next section) and to keep the exponential dependence in $T$ as simple as possible. The radial component of the tetrad-frame energy-momentum tensor, $T_{00}=T_{11}=\Phi$, then reduces to Eq.~(\ref{eq:ikasner_Phi}).

\subsection{\label{subsec:interpretation}Interpretation}

\begin{figure}
    \includegraphics[width=\columnwidth]{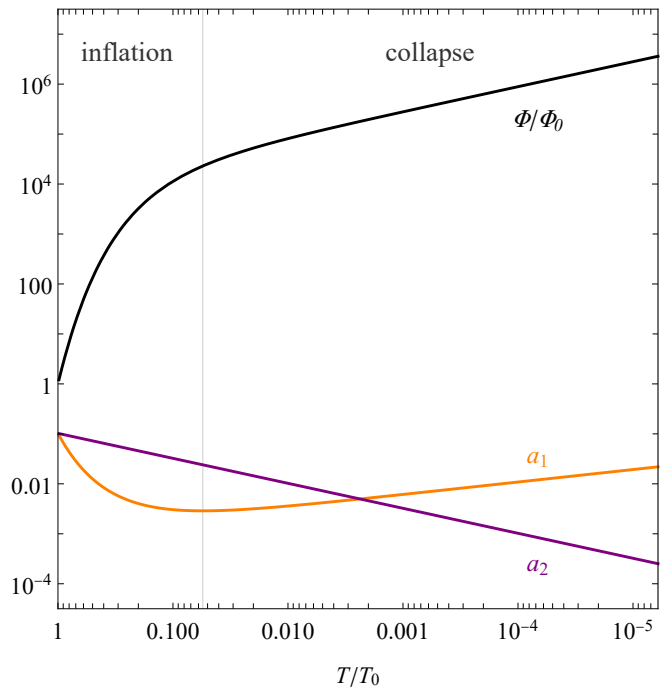}
    \caption{\label{fig:ikasner_evolution} (Color online). Evolution of the inflationary Kasner geometry and radial energy-momentum from the initial Kerr vacuum at time ${T=T_0}$ through inflation and collapse. The plotted quantities are the normalized tetrad-frame mass-energy density $\Phi/\Phi_0$ (black) from Eq.~(\ref{eq:ikasner_Phi}) and the scale factors ${a_1}$ (orange) and ${a_2}$ (purple) from Eq.~(\ref{eq:ikasner_scalefactors}). The parameters chosen here are ${T_0=9.09}$ and ${\Phi_0=0.209}$, so as to match the parameters used in the remaining plots via Eqs.~(\ref{eq:ikasner-cskerr}).}
\end{figure}

The evolution of the inflationary Kasner geometry is visualized in Fig.~\ref{fig:ikasner_evolution}. The evolution begins at ${T=T_0}$, when the mass-energy density $\Phi$ is at its small initial value of $\Phi_0$. At this point, an observer might see such a geometry if, for example, they fall inside a rotating, accreting black hole and approach its inner horizon. Once the observer has come close enough to the inner horizon, the inflation epoch will begin, characterized by the rapid exponentiation of the observed stream energy density. As the observer's proper time progresses forward and $T$ progresses backward from $T_0$, the inflation will slow until $a_1$ turns around (at the vertical gray line in Fig.~\ref{fig:ikasner_evolution}), signaling the start of the collapse epoch. The inflation-collapse transition occurs when $T$ is of order unity, or more precisely, when $H_1$ changes sign from positive to negative at ${T=1/2}$ (independent of the value of ${T_0}$). During the collapse epoch, $\Phi$ continues to increase as the spacetime collapses in the $y$- and $z$-directions and the observer approaches the inflationary Kasner singularity at ${T=0}$ where the inner horizon would have been.

We have chosen to call the metric of Eq.~(\ref{eq:ikasner}) the inflationary Kasner metric because of its similarity to a homogeneous vacuum solution first found by the mathematician Edward Kasner in 1921 \cite{kas21}. For three spatial dimensions, the Kasner metric has the line element
\begin{equation}
    ds^2=-dT^2+a_1^2dx^2+a_2^2dy^2+a_3^2dz^2,
\end{equation}
where the scale factors $a_i$ evolve purely as power laws with $T$,
\begin{equation}
    a_i=T^{p_i},
\end{equation}
for exponents $p_i$ that were found in the vacuum solution to satisfy the following conditions:
\begin{equation}
    \sum_ip_i=\sum_ip_i^2=1.\label{eq:kasner_conditions}
\end{equation}

From these Kasner conditions it can be shown that one of the exponents must always be negative or zero while the other two are nonnegative. More specifically, if the Kasner exponents are labelled in increasing order, they will satisfy the condition
\begin{equation}
    -\frac{1}{3}\le p_1\le0\le p_2\le\frac{2}{3}\le p_3\le1.\label{eq:kasner_exponents}
\end{equation}

This mathematical picture can be physically interpreted as an evolution in which one spacetime axis expands while the other two collapse (assuming the time coordinate $T$ is positive and decreases with increasing proper time; otherwise the Kasner solution would describe a globally expanding spacetime).

The significance and applicability of the Kasner metric for black hole interiors was explored in the 1970s by Belinskii, Khalatnikov, and Lifschitz, who described a collapse consisting of a series of ``Kasner epochs'' during which the geometry is approximated by a Kasner metric with constant Kasner exponents $p_i$ \cite{bel70}. According to the BKL model, the three spatial components of the metric evolve in such a way so that the metric determinant decreases monotonically to zero in a finite time, but one spatial component always increases while the other two decrease (cf.~Eq.~(\ref{eq:kasner_exponents})). Once one of the decreasing components has collapsed to a small enough value, the geometry then undergoes a ``BKL bounce,'' in which one of the two collapsing components begins to grow, the previously expanding component begins to collapse, and the angles of orientation for the collapsing and expanding axes change.

In 2017, the evolution of the inner horizon of a rotating, accreting black hole was explored numerically by Hamilton, who found that the spacetime approximately undergoes a BKL collapse as predicted a half a century earlier \cite{ham17}. Here, we find that the inflationary Kasner solution is an analytic model of such a collapse, with two Kasner epochs as described below.

The first epoch in the inflationary Kasner solution, labelled ``inflation'' on Fig.~\ref{fig:ikasner_evolution}, begins at ${T=T_0}$. During inflation, the exponential terms in the line element of Eq.~(\ref{eq:ikasner}) dominate the evolution of the geometry, so that the scale factor for the $x$-axis collapses while those of the $y$- and $z$-axes remain approximately static. The behavior can thus be approximated as that of Minkowski space with accelerated radar-like coordinates in the $x$-direction \cite{min05}. The inflation epoch resembles a Kasner epoch with exponents
\begin{equation}
    (p_1,p_2,p_3)=(1,0,0),
\end{equation}
corresponding to a spacetime collapsing only in the radial direction (indeed, the growing streams focus along the principal null directions during inflation). The inflation continues as the locally-measured energy-momentum $\Phi$ grows at an absurdly fast rate with a scale factor of order ${\sim\text{e}^{1/\Phi_0}}$ (as confirmed in the next section, in which it is found that the Kasner time $T_0$ scales as ${1/u\sim1/\Phi_0}$). For an astronomically realistic black hole, in which the initial mass-energy density of accreted matter or radiation is generally quite small after the initial collapse, $\Phi$ could reach 10$^{100}$ and beyond. It is perhaps fitting that Kasner himself (with his nine-year-old nephew) was the coiner of the term ``googol'' \cite{kas40}.

Once the Kasner time $T$ has grown small enough, the exponential terms in the Eq.~(\ref{eq:ikasner}) freeze out, leaving the power laws in $T$ to dominate the geometry's evolution. The result is the collapse epoch, beginning at around ${T=1/2}$, in which the scale factor for the $x$-axis turns around and begins to grow, the scale factors in the $y$- and $z$-directions continue to collapse, and the streams' energy-momenta continues to grow, albeit at a slower rate in $\log(T)$. This corresponds to a Kasner epoch with exponents
\begin{equation}
    (p_1,p_2,p_3)=\left(-\frac{1}{3},\frac{2}{3},\frac{2}{3}\right),
\end{equation}
which can be found by a coordinate transformation of $T$ from the ${\left(-\frac{1}{4},\frac{1}{2},\frac{1}{2}\right)}$ form of Eq.~(\ref{eq:ikasner}) in order to satisfy the Kasner conditions of Eq.~(\ref{eq:kasner_conditions}).  This epoch approximates a Schwarzschild geometry asymptotically close to the Schwarzschild singularity. To see why this is the case, note that in the limit as ${r\to0}$, the Schwarzschild line element takes the form
\begin{equation}
    ds^2\approx\frac{2M}{r}dt^2-\frac{r}{2M}dr^2+r^2do^2,
\end{equation}
where ${do^2=d\theta^2+\sin^2\!\theta\ d\phi^2}$ is the 2-sphere line element.

With the coordinate transformations ${r\to T^{2/3}}$ and ${t\to x}$ (note that $r$ is timelike and $t$ spacelike in this regime), the line element becomes
\begin{equation}
    ds^2\approx-dT^2+T^{-2/3}dx^2+T^{4/3}do^2,
\end{equation}
where the constants have been absorbed into the coordinates for simplicity. This is precisely the ${\left(-\frac{1}{3},\frac{2}{3},\frac{2}{3}\right)}$ Kasner epoch when the $\theta$-$\phi$ plane is transformed into the $y$-$z$ plane.

Thus, the inflationary Kasner metric provides a simple model that encompasses all the relevant features of the evolution of the geometry near the inner horizon of a rotating, accreting black hole as it undergoes a BKL-like collapse. That collapse consists of two Kasner epochs, an inflationary epoch characterized by Kasner exponents ${(1,0,0)}$ that matches the behavior of the traditional Poisson-Israel mass-inflation regime, and a subsequent collapse epoch characterized by Kasner exponents ${\left(-\frac{1}{3},\frac{2}{3},\frac{2}{3}\right)}$ as the geometry approaches a spacelike singularity at ${T=0}$. 

In the next section, we confirm the applicability of this model to astrophysical inner horizons by comparing it to a more complex model, the conformally-separable Kerr model, with the eventual goal of finding the necessary boundary conditions to attach the inflationary Kasner metric to the Kerr metric far enough above the inner horizon.

\section{\label{sec:application}Matching near the inner horizon}

We have yet to verify explicitly that the assumptions of the inflationary Kasner metric hold true near the inner horizon of an astrophysical black hole. In order to do so, we employ Hamilton's conformally-separable Kerr metric, which has already been shown to provide a reasonable classical model of the inner workings of an accreting black hole \cite{ham11a,ham11b,ham11c}.

In Sec.~\ref{subsec:cskerr}, we review the conformally-separable model, finding that it exactly matches the behavior of the inflationary Kasner model for asymptotically small accretion rates. Specifically, we find in Eqs.~(\ref{eq:ikasner-cskerr}) a set of transformations between the parameters and coordinates of the inflationary Kasner and conformally-separable Kerr models. These relations confirm the validity and applicability of the inflationary Kasner model to an astrophysical inner horizon.

Then, in Sec.~\ref{subsec:geodesics}, we use the transformations of Eqs.~(\ref{eq:ikasner-cskerr}) to match the inflationary Kasner solution to the Kerr metric. Such a matching allows us to ray-trace null geodesics across both regimes, from the Kerr background to an inflationary Kasner observer.

\subsection{\label{subsec:cskerr}Conformally-separable Kerr solution}

Just as the inflationary Kasner metric provides a non-vacuum generalization of the vacuum solution of the Kasner metric to include the effects of accretion, so too does the conformally-separable Kerr metric provide a generalization of the vacuum Kerr solution to include the effects of accretion. Here we present the main results of the conformally-separable model; a more complete review can be found in Refs.~\cite{ham11a,ham11b,ham11c} (or, in the Boyer-Lindquist form used here, in Ref.~\cite{ham17}).

Consider a rotating, accreting black hole with external mass $M$. For an ideal, rotating Kerr black hole, three assumptions hold true: the black hole is axisymmetric, the spacetime is stationary, and its Hamilton-Jacobi equations are separable \cite{car68}. The conformally-separable model presented below was developed in an attempt to find the most general metric that still satisfies these conditions. To allow for the inclusion of accreting matter or radiation, however, the conditions required slight modification. In particular, instead of strict stationarity, the assumption of conformal stationarity adopted here implies that the spacetime expands in a self-similar fashion with time at an asymptotically small rate (this rate is the accretion rate $v$ that is taken to be asymptotically close to zero in Eq.~(\ref{eq:cskerr_T11_smallv})) \cite{ham11b}. Such a condition may not apply at the onset of a gravitational collapse when the accretion is supplied by the bulk of the collapsing matter, but that collapse occurs within a small proper time, and at late times, a black hole will only grow at a rate on the order of its light crossing time divided by the age of the Universe, a very small number. However, it will still accrete, so the assumptions of isolation and Price tail decay from models with a null weak singularity at the inner horizon will not apply.

It should be noted that strictly speaking, there is no inner horizon in the conformally-separable model (nor in the inflationary Kasner model), since mass inflation near that region of spacetime will give way to collapse. When we refer to the inner horizon, we thus mean the region of spacetime within the black hole asymptotically close to the dimensionless Boyer-Lindquist radius ${r_-\equiv 1-\sqrt{1-a^2}}$, in which crossing streams focus along the principal null directions and cause inflation and collapse. Also, strictly speaking, the conformally-separable model does not hold for extremal black holes, for which $\Delta_0'$ defined in Eq.~(\ref{eq:cskerr_deltaprime}) is zero. However, this should not be too worrisome, since astronomically realistic black holes are expected to have spins no higher than the Thorne limit \cite{tho74}.

Under the assumptions of conformal stationarity, axial symmetry, and conformal separability, the conformally-separable line element takes the form
\begin{align}
    ds^2&=\rho_s^2\text{ e}^{2(vt-\xi)}\bigg(\frac{dr^2}{\left(r^2+a^2\right)^2\text{e}^{3\xi}\Delta_r}+\frac{\sin^2\!\theta}{\Delta_\theta}d\theta^2\nonumber\\
    &+\frac{-\text{e}^{3\xi}\Delta_r(dt-\omega_\theta d\phi)^2+\Delta_\theta(d\phi-\omega_rdt)^2}{(1-\omega_r\omega_\theta)^2}\bigg)\label{eq:cskerr}
\end{align}
(\cite{ham17}), where ${x^\mu=\{r,t,\theta,\phi\}}$ are dimensionless Boyer-Lindquist coordinates (the radial coordinate is written first to emphasize that $r$ is timelike within the outer horizon). The function $\Delta_r$ is the horizon function, whose zeros define the location of the geometry's horizons, and $\Delta_\theta$ is the polar function, whose zeros define the location of the north and south poles. Additionally, $\omega_r$ is the angular velocity of the principal frame through the coordinates, and $\omega_\theta$ is the specific angular momentum of principal null congruence photons. The $r$ and $\theta$ subscripts denote functions of only $r$ and $\theta$, respectively, and $\rho_s$ is the separable part of the conformal factor. Eq.~(\ref{eq:cskerr}) reduces to the familiar Kerr line element when the following definitions are made:
\begin{subequations}\label{eq:kerr_limit}
\begin{gather}
    \Delta_r=\frac{r^2-2r+a^2}{\left(r^2+a^2\right)^2},\qquad\Delta_\theta=\sin^2\!\theta,\label{eq:kerr-deltas}\\
    \omega_r=\frac{a}{r^2+a^2},\qquad\omega_\theta=a\sin^2\!\theta,\\
    \rho_s=M\sqrt{r^2+a^2\cos^2\!\theta},\\
    \xi=v=0,\label{eq:kerr-xiv}\\
    \{r,t,\theta,\phi\}=\left\{\frac{r_\text{BL}}{M},\frac{t_\text{BL}}{M},\theta_\text{BL},\phi_\text{BL}\right\},
\end{gather}
\end{subequations}
where $M$ is the black hole's external mass, ${a\equiv J/M^2}$ is the black hole's dimensionless spin parameter, and ${\{r_\text{BL},t_\text{BL},\theta_\text{BL},\phi_\text{BL}\}}$ are the standard (dimensionful) Boyer-Lindquist coordinates.

If the vacuum Kerr form of Eq.~(\ref{eq:cskerr}) is generalized to include the effects of accretion, the solution to Einstein's equations sourced by ingoing and outgoing collisionless null streams implies that three of the above definitions in Eqs.~(\ref{eq:kerr_limit}) are amended:

(1) The dimensionless factor $v$ becomes an arbitrary free parameter, which can be interpreted (with the proper gauge choice) as the black hole's net accretion rate $\dot{M}$, or equivalently, the difference in the flux of outgoing and ingoing streams near the inner horizon. This factor can be treated as very small and reduces to zero for equal streams of ingoing and outgoing radiation.

(2) The inflationary exponent $\xi$, which measures the degree to which the geometry has undergone self-similar compression, changes with the radius and accretion parameters, behaving like a step function near the inner horizon as inflation is ignited.

(3) The horizon function $\Delta_r$ strays from its Kerr value near the inner horizon, ``freezing out'' at a small, negative value during collapse instead of reaching zero at ${r=r_-}$.

In the conformally-separable solution, $\xi$ and $\Delta_r$ are governed by the highly nonlinear pair of relations in Eq.~88 of Ref.~\cite{ham11b} (where ${x=\frac{1}{a}\cot^{-1}\!\left(\frac{r}{a}\right)}$, ${y=-\cos\theta}$, and ${\Delta_x=\text{e}^{3\xi}\Delta_r}$). To simplify their behavior, it suffices to assume their Kerr values (Eqs.~(\ref{eq:kerr-deltas}) and (\ref{eq:kerr-xiv})) for all portions of spacetime except just above the inner horizon. In the regime near the inner horizon, $\xi$ rapidly increases from zero as $r$ remains frozen at its inner horizon value of $r_-$, and the equations governing the evolution of $\xi$ and $\Delta_r$ simplify to
\begin{align}
    &\text{e}^{\xi}=\left(\frac{(U_r+v)(U_r-v)}{(u+v)(u-v)}\right)^{1/4},\label{eq:cskerr_xiU}\\
    \Delta_r&=\Delta_0\left(\frac{(U_r+v)(u-v)}{(U_r-v)(u+v)}\right)^{\Delta'/(4v)},\label{eq:cskerr_deltar}
\end{align}
where $\Delta_0$ is a constant of integration equal to the linear extrapolation of $\Delta_r$ evaluated away from the inner horizon when ${\xi=0}$, ${U_r=u}$, and $\Delta_r$ still equals its Kerr value. The dimensionless parameter $u$, the counter-streaming velocity, represents the average of the initial accretion rates from the two streams. The accretion parameters satisfy ${0<v<u\ll1}$, and the outgoing and ingoing accretion rates are proportional to ${u\pm v}$. The function $U_r$ is defined by
\begin{equation}
    U_r\equiv\frac{d\xi}{dr}(r^2+a^2)\text{ e}^{3\xi}\Delta_r.
\end{equation}

In addition, the constant $\Delta'$ in Eq.~(\ref{eq:cskerr_deltar}) is proportional to the radial derivative of the Kerr horizon function evaluated at the inner horizon:
\begin{align}\label{eq:cskerr_deltaprime}
\begin{split}
    \Delta'&\equiv-\frac{d\Delta_r}{dr}\bigg|_{r_-}(r_-^2+a^2)\\
    &=\frac{2\left(r_-^3-3r_-^2+a^2r_-+a^2\right)}{(r_-^2+a^2)^2}.
\end{split}
\end{align}

\begin{figure}
    \includegraphics[width=\columnwidth]{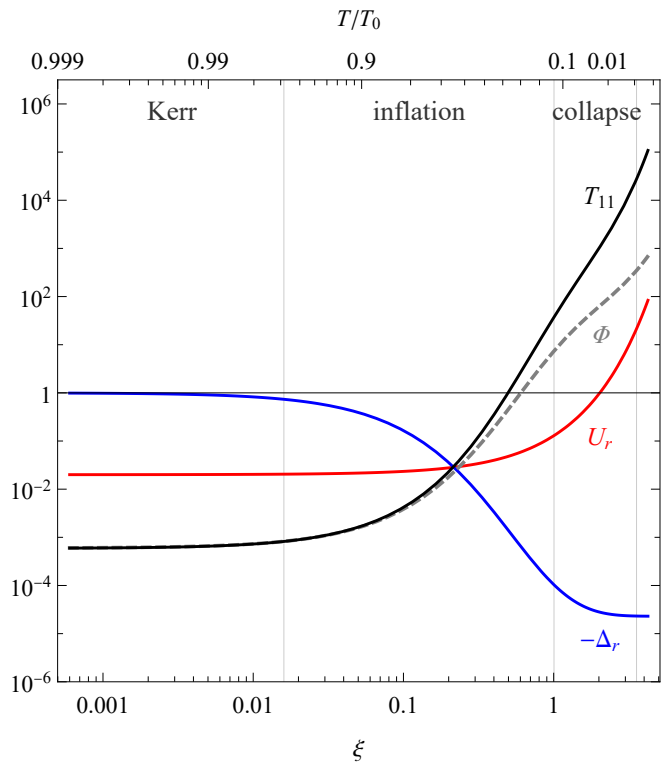}
    \caption{\label{fig:cskerr_evolution} (Color online). Evolution of quantities in the conformally-separable Kerr model. Plotted are the tetrad-frame energy-momentum component $T_{11}$ from Eq.~(\ref{eq:cskerr_T11}) (black), the corresponding inflationary Kasner energy-momentum $\Phi$ (the ${v\to0}$ limit of the conformally-separable $T_{11}$) (gray dashed), the parameter $U_r$ from Eq.~(\ref{eq:cskerr_xiU}) (red), and the magnitude of the horizon function $\Delta_r$ from Eq.~(\ref{eq:cskerr_deltar}) (blue). The parameters of this model have been chosen to avoid numerical overflow while still allowing the solution to capture the full behavior; in particular, ${u=0.02}$, ${v=0.01}$, and ${a=0.96}$. The difference in the appearance of $\Phi$ here vis-\`a-vis Fig.~\ref{fig:ikasner_evolution} is solely due to a difference in the scaling of the axes.}
\end{figure}

The conformally-separable Kerr model predicts that the geometry of the inner horizon will be divided into three distinct epochs, as shown in Fig.~\ref{fig:cskerr_evolution}. The parameter $\xi$ represents the timelike coordinate separating these epochs, just as $T$ does for the inflationary Kasner model (in fact, it will be shown later, Eqs.~(\ref{eq:ikasner-cskerr}), that the identification ${T\propto\text{e}^{-2\xi}}$ generally holds).

Initially, the geometry resembles the Kerr vacuum when $\xi$ is negligibly small and $r$ is just above its inner horizon value. Then, as $r$ approaches $r_-$, the mass inflation epoch begins as the locally-measured radial energy-momentum $T_{11}$ of the streams rapidly inflates (along with the internal mass parameter and the Weyl curvature). The horizon function dominates the geometry's evolution during this epoch as $\Delta_r$ deflates toward $0^-$. Throughout the inflation and collapse epochs, $r$ remains approximately frozen at its inner horizon value of $r_-$.

Finally, inflation is slowed once $\xi$ grows large enough and begins to dominate, causing a self-similar collapse of the geometry. During the collapse epoch, the curvature and $T_{11}$ once again begin to diverge, while the horizon function freezes out at an exponentially small value. The collapse epoch then continues until Eqs.~(\ref{eq:cskerr_xiU}) and (\ref{eq:cskerr_deltar}) are no longer valid because of the increasing angular motion of the streams. However, it is possible that the conformally-separable solution will break down regardless after this point, once the curvature exceeds the Planck scale.

The connections between the conformally-separable Kerr model and the inflationary Kasner model become evident when considering the energy-momentum tensor seen in a tetrad frame. A natural tetrad frame to choose is the one encoded by the line element in Eq.~(\ref{eq:cskerr}), with 1-forms
\begin{subequations}\label{eq:cartertetrad}
\begin{align}
    e^0_{\ \mu}dx^\mu&=\frac{\rho_s\text{ e}^{vt-5\xi/2}}{(r^2+a^2)\sqrt{-\Delta_r}}dr,\\
    e^1_{\ \mu}dx^\mu&=\frac{\rho_s\text{ e}^{vt+\xi/2}\sqrt{-\Delta_r}}{1-\omega_r\omega_\theta}\left(dt-\omega_\theta d\phi\right),\\
    e^2_{\ \mu}dx^\mu&=\frac{\rho_s\text{ e}^{vt-\xi}\sin\theta}{\sqrt{\Delta_\theta}}d\theta,\\
    e^3_{\ \mu}dx^\mu&=\frac{\rho_s\text{ e}^{vt-\xi}\sqrt{\Delta_\theta}}{1-\omega_r\omega_\theta}\left(d\phi-\omega_rdt\right).
\end{align}
\end{subequations}

In the Kerr limit, the tetrad frame in Eqs.~(\ref{eq:cartertetrad}) reduces to the interior Carter frame, in which observers at rest see the principal null directions as purely radial (in the $x^1$-direction) as the frame follows them freely falling and rotating inward. The interior Carter frame differs from the standard (exterior) Carter frame only in the swapping of ${e^0_{\ \mu}\leftrightarrow e^1_{\ \mu}}$ and ${\sqrt{-\Delta_r}\leftrightarrow\sqrt{+\Delta_r}}$, since below the outer horizon, $r$ becomes timelike and ${\Delta_r}$ becomes negative.

In this tetrad frame, Einstein's equations yield the following non-negligible components of the energy-momentum tensor seen by a Carter observer:
\begin{equation}\label{eq:cskerr_T11}
    T_{00}=T_{11}=\frac{U_r\Delta'-v^2}{8\pi\rho_s^2\text{ e}^{2vt+\xi}(-\Delta_r)}
\end{equation}
(cf.~Eqs.~125-128 in Ref.~\cite{ham11b}). These components, which rapidly diverge during inflation and collapse (see Fig.~\ref{fig:cskerr_evolution}), represent the net combination of the energy-momenta of ingoing and outgoing collisionless streams observed in the radial direction. Their behavior is dominated by the vanishing of $\Delta_r$ during inflation and by the conformal piece $\text{e}^{-\xi}$ once the horizon function freezes out during collapse. In terms of the counter-streaming velocity $u$, a Taylor expansion for small $v$ yields
\begin{equation}\label{eq:cskerr_T11_smallv}
    T_{00}=T_{11}\approx\frac{\text{e}^{\xi}u\Delta'}{8\pi\rho_s^2(-\Delta_0)}\text{ e}^{\frac{\Delta'}{2u}\left(1-\text{ e}^{-2\xi}\right)}+\mathcal{O}\left(v\right).
\end{equation}

The radial energy-momentum thus grows as ${\sim\text{e}^{1/u}}$, so that, perhaps counterintuitively, the smaller the value of $u$, the more rapid the inflation. For astronomically realistic black holes, the above expansion is generally valid, since $v$ scales as the black hole light crossing time ${t_\text{BH}}$ divided by the accretion (mass-doubling) time ${t_\text{acc}}$, and for most of the lifetime of the black hole, ${t_\text{acc}\gg t_\text{BH}}$ \cite{ham10}.

When comparing the energy-momentum tensor of the conformally-separable Kerr metric in Eq.~(\ref{eq:cskerr_T11_smallv}) with the energy-momentum tensor of the inflationary Kasner metric in Eq.~(\ref{eq:ikasner_Phi}), the two are equivalent in the limit ${v\to0}$, when the following definitions are made:
\begin{subequations}\label{eq:ikasner-cskerr}
\begin{align}
    T&=T_0\text{ e}^{-2\xi},\\
    T_0&=\frac{\Delta'}{2u},\\
    \Phi_0&=\frac{u\Delta'}{8\pi\rho_-^2(-\Delta_0)},
\end{align}
\end{subequations}
where $\rho_-$ is the value of the separable conformal factor $\rho_s$ at the inner horizon.

Thus, the inflationary Kasner solution provides a simple yet precise approximation of the conformally-separable Kerr spacetime seen in the tetrad rest frame of a Carter observer, through the matching of Eqs.~(\ref{eq:ikasner-cskerr}). The conformally-separable solution, in turn, provides an approximation of the geometry of a rotating, accreting black hole, which reduces to the inflationary Kasner solution near the inner horizon in the limit of an asymptotically small accretion rate $v$.

$T_{00}$ and $T_{11}$ remain the only non-negligible components of $T_{\hat{m}\hat{n}}$ through inflation and collapse, and the collapse epoch of the conformally-separable solution is defined to end when other components of $T_{\hat{m}\hat{n}}$ (namely, $T_{12}$ and $T_{22}$), which initially diverge at a much slower rate, become comparable in magnitude to the radial components. This occurs at ${\xi=\Delta'/(6u)-\ln\sqrt{-\Delta_0}}$ (or equivalently, at ${T/T_0=-\Delta_0\text{ e}^{-\Delta'/(3u)}}\approx10^{-5}$ in Fig.~\ref{fig:ikasner_evolution}), and beyond this point, the approximations of Eqs.~(\ref{eq:cskerr_xiU}), (\ref{eq:cskerr_deltar}), and (\ref{eq:cskerr_T11}) are no longer valid. The classical solution can be continued numerically for higher $\xi$, yielding a series of even more complex Kasner epochs and BKL bounces, although an extension of the classical solution may fail if quantum effects become important once the curvature passes the Planck scale \cite{ham17}.

\subsection{\label{subsec:geodesics}Null geodesic behavior}

What will an observer in the inflationary Kasner spacetime see? To answer this question, consider a Carter observer (at rest in the tetrad of Eqs.~(\ref{eq:cartertetrad})) falling into a rotating black hole from rest at infinity and approaching the inner horizon.

The Kerr metric provides an excellent approximation of a rotating black hole's geometry far above the inner horizon, so the Kerr null geodesic equations will provide the trajectory of a photon in this regime. However, once the observer approaches the inner horizon, streams of ingoing and outgoing matter will focus along the radial directions in the Carter tetrad and will begin to inflate, causing the geometry to be better approximated by the inflationary Kasner metric. Thus, here we find the equations for null geodesics in the inflationary Kasner spacetime and then connect them to null geodesics in the Kerr spacetime in a regime near the inner horizon where both are valid.

To find null geodesic trajectories in the inflationary Kasner spacetime, note that because the metric is homogeneous, there are three conserved quantities corresponding to each of the spatial coordinates $x$, $y$, and $z$. These momenta are simply the covariant forms of the spatial components of a photon's coordinate-frame four-momentum,
\begin{equation}\label{eq:ikasner_ki}
    k_i=g_{i\mu}\frac{dx^\mu}{d\lambda},\qquad\text{where }i\in\{x,y,z\}.
\end{equation}

When Eqs.~(\ref{eq:ikasner_ki}) are combined with the condition ${k^\mu k_\mu=0}$, the four components of the four-momentum can be expressed in terms of the coordinate time $T$ and conserved quantities $k_x$, $k_y$, and $k_z$. In the tetrad frame of Eq.~(\ref{eq:ikasner}), these components take the form
\begin{subequations}\label{eq:k_IK}
\begin{align}
    (k_\text{IK})^0&=-\sqrt{\frac{k_x^2}{a_1^2}+\frac{k_y^2+k_z^2}{a_2^2}},\\
    (k_\text{IK})^1&=\frac{k_x}{a_1},\\
    (k_\text{IK})^2&=\frac{k_y}{a_2},\\
    (k_\text{IK})^3&=\frac{k_z}{a_2},
\end{align}
\end{subequations}
where the subscript $\text{IK}$ denotes quantities valid in the inflationary Kasner regime. The negative sign for ${(k_\text{IK})^0}$ is chosen so that the affine parameter increases as $T$ decreases from $T_0$ just above the inner horizon to $0$ at the inflationary Kasner singularity.

In the coordinate frame, Eqs.~(\ref{eq:k_IK}) lead to the following equations of motion that can be integrated:
\begin{subequations}\label{eq:ikasner_EOM}
\begin{align}
    \frac{dT}{d\lambda}&=-\frac{\mathcal{E}}{a_1}\sqrt{\left(\frac{a_1^{\text{obs}}}{a_1}\right)^2\cos^2\!\chi+\left(\frac{a_2^{\text{obs}}}{a_2}\right)^2\sin^2\!\chi},\\
    \frac{dx}{d\lambda}&=-\frac{\mathcal{E}}{a_1}\frac{a_1^{\text{obs}}}{a_1}\cos\chi,\\
    \frac{dy}{d\lambda}&=-\frac{\mathcal{E}}{a_2}\frac{a_2^{\text{obs}}}{a_2}\sin\chi\cos\psi,\\
    \frac{dz}{d\lambda}&=-\frac{\mathcal{E}}{a_2}\frac{a_2^{\text{obs}}}{a_2}\sin\chi\sin\psi,
\end{align}
\end{subequations}
where $a_i^{\text{obs}}$ is the value of $a_i$ at the observer's position, and the constants of motion $k_i$ have been replaced by the observer's viewing angles ${\chi\in[0,\pi]}$ and ${\psi\in[0,2\pi)}$ (and the normalization factor $\mathcal{E}$), which indicate the position of the photon in the observer's field of view. More details about the definitions of these angles and their relations to other quantities used throughout the paper are given in the \hyperref[sec:appendix]{Appendix}. The important point to note here is that ${\chi=0^{\circ}}$ corresponds to an ingoing photon reaching an observer looking in the principal null direction away from the black hole, and ${\chi=180^{\circ}}$ corresponds to an observer looking directly toward the black hole in the principal null direction.

\begin{figure*}
    \centering
    \begin{minipage}{0.95\textwidth}
        \includegraphics[width=0.95\columnwidth]{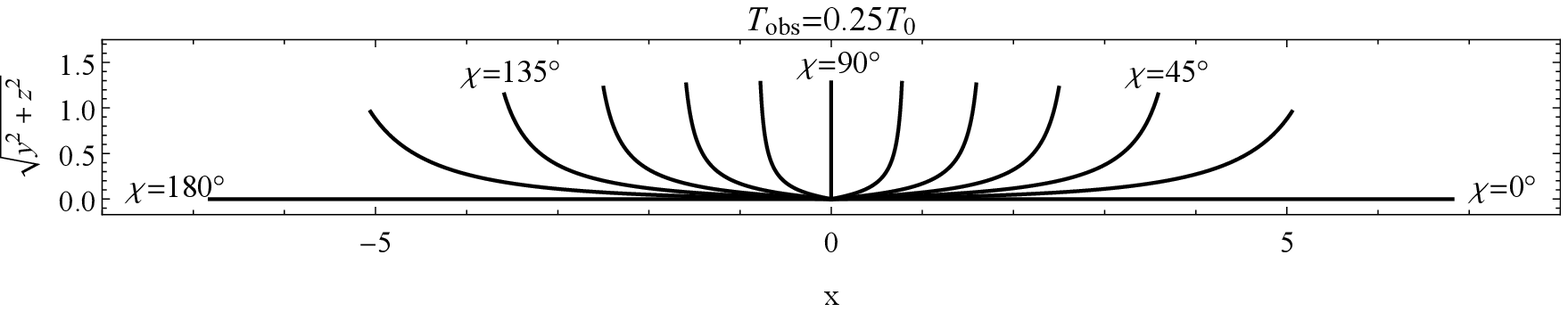}
    \end{minipage}\vspace{8pt}
    \begin{minipage}{0.95\textwidth}
        \includegraphics[width=0.95\columnwidth]{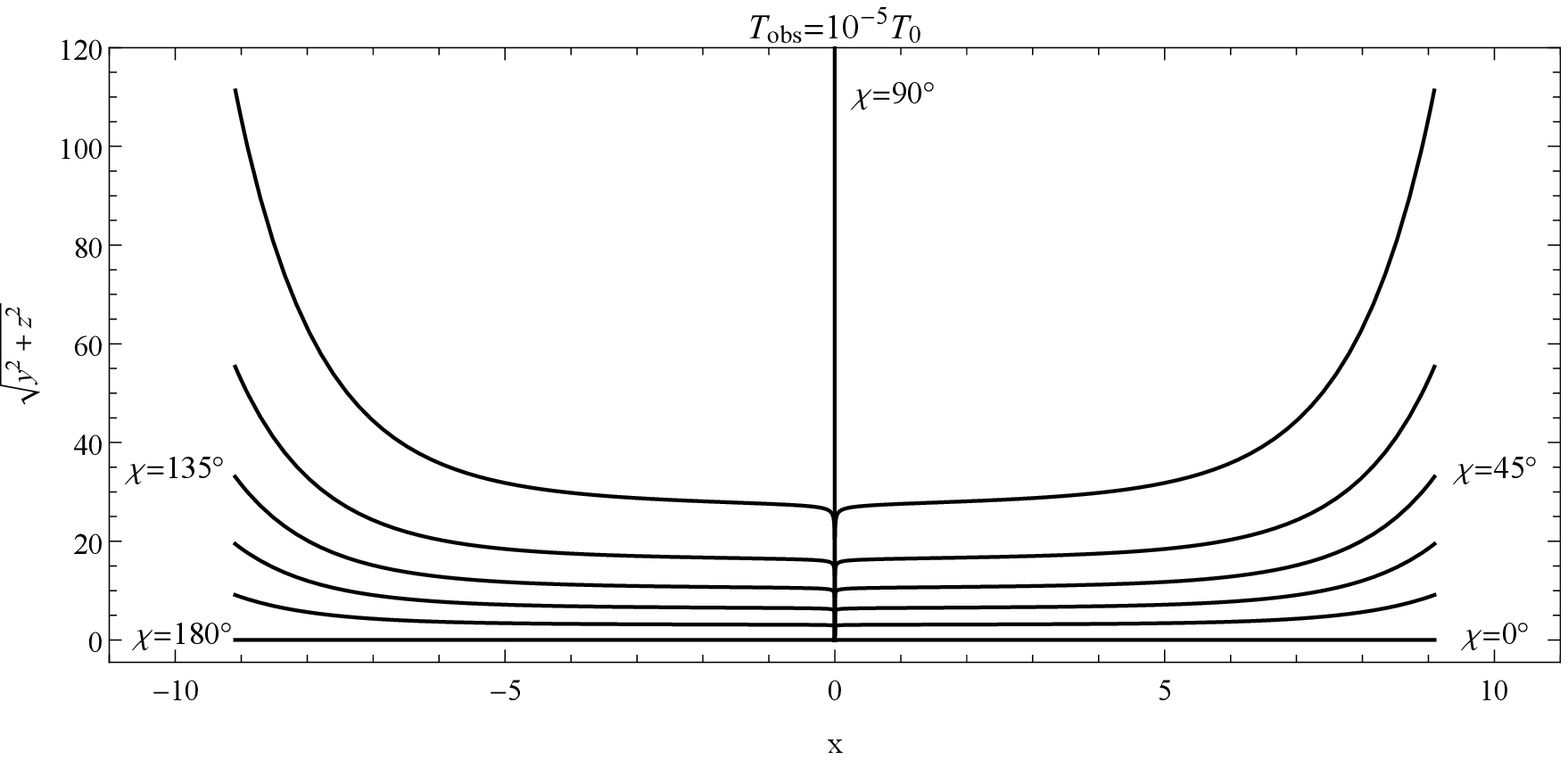}
    \end{minipage}%
    \caption{\label{fig:ikasner_geodesics}Null geodesics seen by an observer at the origin in the inflationary Kasner spacetime. Geodesics are parametrized by the observer's viewing angle $\chi$ at an equal spacing of $15^{\circ}$, from an observer looking directly outward (${\chi=0^{\circ}}$) to one looking directly inward toward the black hole's center (${\chi=180^{\circ}}$). All geodesics end at the origin at ${T=T_{\text{obs}}}$ and are ray-traced backwards via Eqs.~(\ref{eq:ikasner_EOM}) to ${T=T_0}$. The upper plot shows an observer at ${T_{\text{obs}}=0.25T_0}$ near the end of the inflation epoch, and the lower plot shows an observer at ${T_{\text{obs}}=10^{-5}T_0}$ near the end of the collapse epoch. The parameters chosen here are ${T_0=9.09}$ and ${\Phi_0=0.209}$.}
\end{figure*}

The evolution of null geodesics seen by an observer at $T_{\text{obs}}$ looking in different directions is shown in Fig.~\ref{fig:ikasner_geodesics} during both the inflation and collapse epochs. In these plots, the positive $x$-direction is aligned with the principal null direction away from the black hole. Since the inflationary Kasner metric is isotropic in the $y$-$z$ plane, the dependence on the viewing angle $\psi$ is trivial---the geodesics of Fig.~\ref{fig:ikasner_geodesics} can be revolved around the $x$-axis in 3D space to obtain solutions with different values of $\psi$.

The inflation epoch is characterized by the focusing of null geodesics along the principal null directions. An observer in the inflation epoch (upper panel of Fig.~\ref{fig:ikasner_geodesics}) will thus see both ingoing and outgoing null geodesics that have begun to align along the $x$-axis, so that an increasingly large portion of the observer's sky is taken up by a narrowing band of the inflationary Kasner background orthogonal to the principal null axis. The same inflation power law behavior from the upper panel of Fig.~\ref{fig:ikasner_geodesics} is also seen in the lower panel, in which the photons undergo both inflation and collapse. These photons all start at ${T=T_0}$, corresponding to ${x=\pm T_0}$ for all but the ${\chi=90^{\circ}}$ geodesic (which begins at approximately ${\sqrt{y^2+z^2}\propto(T_0/T_\text{obs})^{3/4}}$, far outside the range of this plot). The photons in this plot begin by proceeding inward toward the origin, curving toward the $x$-axis as they undergo inflation. Then, once the photons reach the collapse epoch, they turn sharply, orthogonal to the $x$-axis, until they reach the observer at the origin. As the observer continues farther into the collapse epoch, the turns sharpen even more, and the locations of the turns spread out farther in the ${y-z}$ plane as most of the background radiation from $T_0$ reaching the observer becomes squeezed into a band around ${\chi=90^{\circ}}$. Once the observer has reached the singularity at ${T=0}$ in the Carter tetrad frame, the entire inflationary Kasner background in the observer's field of view will be squashed into the ring at ${\chi=90^{\circ}}$, and photons arriving at any other position in the sky must have originated from a vanishingly small patch of the background along one of the principal null directions.

From the behavior of the null geodesics in Fig.~\ref{fig:ikasner_geodesics}, one must be careful not to jump too quickly to any conclusions about what an observer near the inner horizon would see, especially since, as we shall see, most of the photons arriving at an observer deep in the collapse epoch tend to align almost exactly with part of the boundary of the black hole's shadow. To be certain about each photon's complete path, we must continue the ray-tracing backwards beyond $T_0$ to ${r\gg r_-}$, where only the Kerr solution is valid. 

In the Kerr spacetime, any geodesic is characterized by three conserved quantities: the energy $E$, angular momentum $L$, and Carter constant $K$, defined by
\begin{equation}\label{eq:kerr_ELK}
    E\equiv-k_t,\quad L\equiv k_\phi,\quad K\equiv k_\theta^2+\frac{(k_\phi+\omega_\theta k_t)^2}{\Delta_\theta},
\end{equation}
where $k_t$, $k_\phi$, and $k_\theta$ are the covariant components of a photon's Kerr coordinate-frame four-momentum. Just as with the inflationary Kasner metric, these conserved quantities lead in a straightforward way to the following four-momentum components in the Carter tetrad frame, defined by ${(k_\text{K})^{\hat{m}}\equiv e^{\hat{m}}_{\hspace{5pt}\mu}(dx^\mu/d\lambda)}$:
\begin{subequations}\label{eq:k_K}
\begin{align}
    (k_\text{K})^0&=\pm\frac{1}{\rho_s}\sqrt{K-\frac{(\omega_rL-E)^2}{\Delta_r}},\\
    (k_\text{K})^1&=\frac{\omega_rL-E}{\rho_s\sqrt{-\Delta_r}},\\
    (k_\text{K})^2&=\pm\frac{1}{\rho_s}\sqrt{K-\frac{(L-\omega_\theta E)^2}{\Delta_\theta}},\\
    (k_\text{K})^3&=\frac{L-\omega_\theta E}{\rho_s\sqrt{\Delta_\theta}},
\end{align}
\end{subequations}
where the subscript $\text{K}$ indicates quantities valid in the Kerr regime.

In the coordinate frame, Eqs.~(\ref{eq:k_K}) lead to the following equations of motion:
\begin{subequations}\label{eq:kerr_EOM}
\begin{align}
    \frac{dr}{d\lambda'}&=\pm\sqrt{R(r)},\\
    \frac{dt}{d\lambda'}&=\frac{E-\omega_rL}{\Delta_r}+\frac{L-\omega_\theta E}{\Delta_\theta}\omega_\theta,\\
    \frac{d\theta}{d\lambda'}&=\pm\sqrt{\Theta(\theta)},\\
    \frac{d\phi}{d\lambda'}&=\frac{E-\omega_rL}{\Delta_r}\omega_r+\frac{L-\omega_\theta E}{\Delta_\theta},
\end{align}
\end{subequations}
written in terms of the Mino time ${d\lambda'\equiv d\lambda/\rho_s^2}$ and the effective potentials
\begin{subequations}
\begin{align}
    R(r)&\equiv\left((\omega_r L-E)^2-K\Delta_r\right)(r^2+a^2)^2,\\
    \Theta(\theta)&\equiv\left(K\Delta_\theta-(L-\omega_\theta E)^2\right)\csc^2\!\theta.
\end{align}
\end{subequations}

Below the outer horizon, ${(k_\text{K})^0}$ must be negative, since the radial coordinate is timelike and decreases as the affine parameter increases, so that all geodesics are necessarily infalling. Thus, only the lower sign for Eq.~(\ref{eq:k_K}a) and Eq.~(\ref{eq:kerr_EOM}a) is valid below the outer horizon. However, ${(k_\text{K})^1}$ may be positive or negative in this regime depending on the relative magnitudes of $L$ and $E$, and a geodesic with positive (negative) ${(k_\text{K})^1}$ is said to be outgoing (ingoing). Additionally, a positive (negative) sign for Eq.~(\ref{eq:k_K}c) and Eq.~(\ref{eq:kerr_EOM}c) corresponds to a geodesic whose polar angle $\theta$ increases (decreases) as the affine parameter increases.

The Kerr and inflationary Kasner metrics are both valid in a small domain just above the inner horizon, and we choose to match their null geodesics at the Boyer-Lindquist radius $r_1$ and corresponding Kasner time $T_1$. The exact value of these parameters is not too important; the results of matching the null geodesics are robust for a range of values as long as $T$ is close enough to $T_0$ that $\Delta_r$ is well-approximated by the Kerr horizon function but far enough into the inflation epoch that $r$ has frozen out and the streams have begun to focus along the principal null directions, so that the inflationary Kasner solution is valid. Practically, for the parameters used in the plots throughout this paper, we choose to match geodesics at ${r_1=0.73}$ (with the inner horizon at $r_-=0.72$), corresponding to ${T_1\approx0.388T_0}$.

The assumption ${(k_\text{IK})^{\hat{m}}|_{T=T_1}=(k_\text{K})^{\hat{m}}|_{r=r_1}}$, matching Eqs.~(\ref{eq:k_IK}) and (\ref{eq:k_K}), leads to a direct mapping between the orbital parameters ${(k_x, k_y, k_z)}$ and ${(E, L, K)}$:
\begin{subequations}\label{eq:ELKofki}
\begin{align}
    E&=\frac{\rho_s}{1-\omega_{r}\omega_{\theta}}\left(\frac{k_z\omega_{r}\sqrt{\Delta_{\theta}}}{a_2}-\frac{k_x\sqrt{-\Delta_{r}}}{a_1}\right),\\
    L&=\frac{\rho_s}{1-\omega_{r}\omega_{\theta}}\left(\frac{k_z\sqrt{\Delta_{\theta}}}{a_2}-\frac{k_x\omega_{\theta}\sqrt{-\Delta_{r}}}{a_1}\right),\\
    K&=\frac{\rho_s^2}{a_2^2}\left(k_y^2+k_z^2\right),
\end{align}
\end{subequations}
where the functions $\Delta_r$, $\Delta_\theta$, $\omega_r$, $\omega_\theta$, $\rho_s$, $a_1$, and $a_2$ are all evaluated at the point of matching just above the inner horizon, where ${T=T_1}$, ${r=r_1}$, and ${\theta=\theta_1}$. Additionally, in order to obtain the complete Kerr solution, the proper signs must be specified. With reference to Eqs.~(\ref{eq:kerr_EOM}), one must require:
\begin{subequations}
\begin{align}
    \text{sgn}\left(\pm\sqrt{R(r)}\right)&=-1,\\
    \text{sgn}\left(\pm\sqrt{\Theta(\theta)}\right)&=\text{sgn}\left(k_y\right).
\end{align}
\end{subequations}

With this matching, it is then possible to continue the inflationary Kasner geodesics of Fig.~\ref{fig:ikasner_geodesics} to their points of origin in the Kerr spacetime. Here we consider two domains for the points of origin of Kerr photons: the first source is the fixed background of stars, galaxies, and radiation travelling inward from infinity, and the second source is the collapsing surface of the star that formed the black hole, emitting radiation outward. By the time photons from the latter source reach the observer, they will be so redshifted and dimmed that the star's surface will be practically imperceptible, so any part of the observer's sky consisting solely of photons from this source will form the black hole's shadow. As an example, the schematic Penrose diagram of Fig.~\ref{fig:penrose} shows the paths of ingoing and outgoing photons from both of these sources reaching an observer near the inner horizon at point O.

It may seem counterintuitive that outward-directed photons from the collapsing star's surface near ${r\approx r_+}$ could reach an observer near ${r_-}$. The paths of these photons fall under two general cases: if the photons were emitted during the collapse just before the formation of the event horizon (at ${r>r_+}$), they may reach a turning point below the photon sphere and travel inward until reaching the observer. Alternatively, if they were emitted below the event horizon (at ${r<r_+}$), they can remain outgoing as their Boyer-Lindquist radius decreases, until they are detected by an observer looking inward.

\begin{figure*}
    \centering
    \begin{minipage}{0.45\textwidth}
        \includegraphics[width=0.95\textwidth]{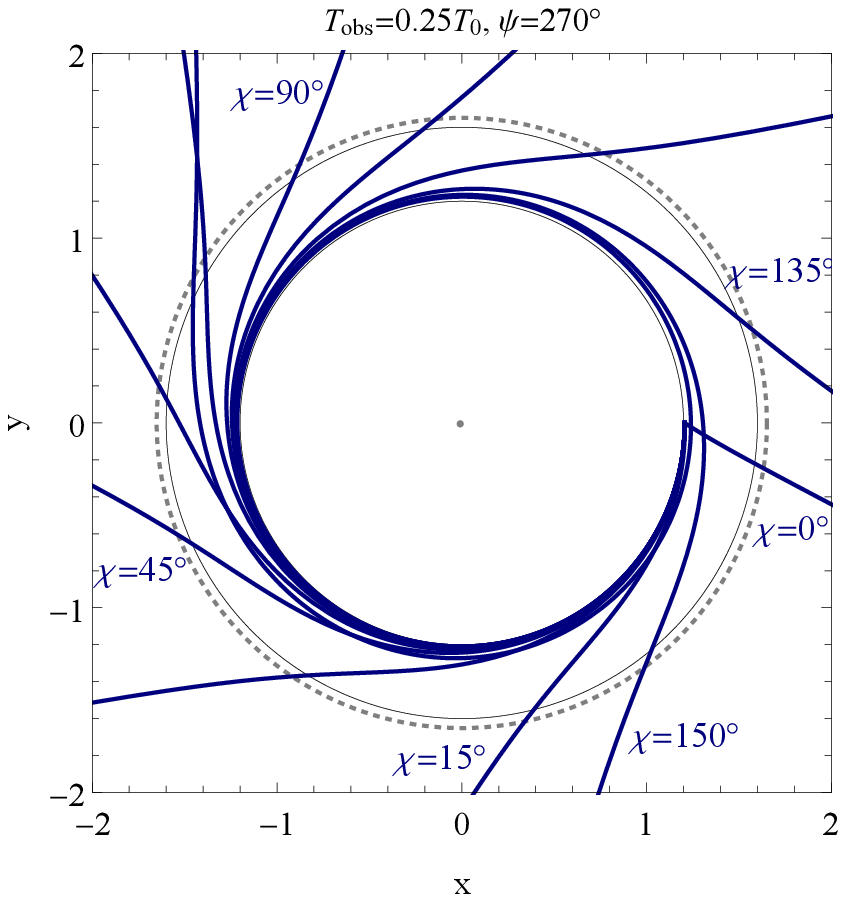}
    \end{minipage}
    \begin{minipage}{0.45\textwidth}
        \includegraphics[width=0.95\textwidth]{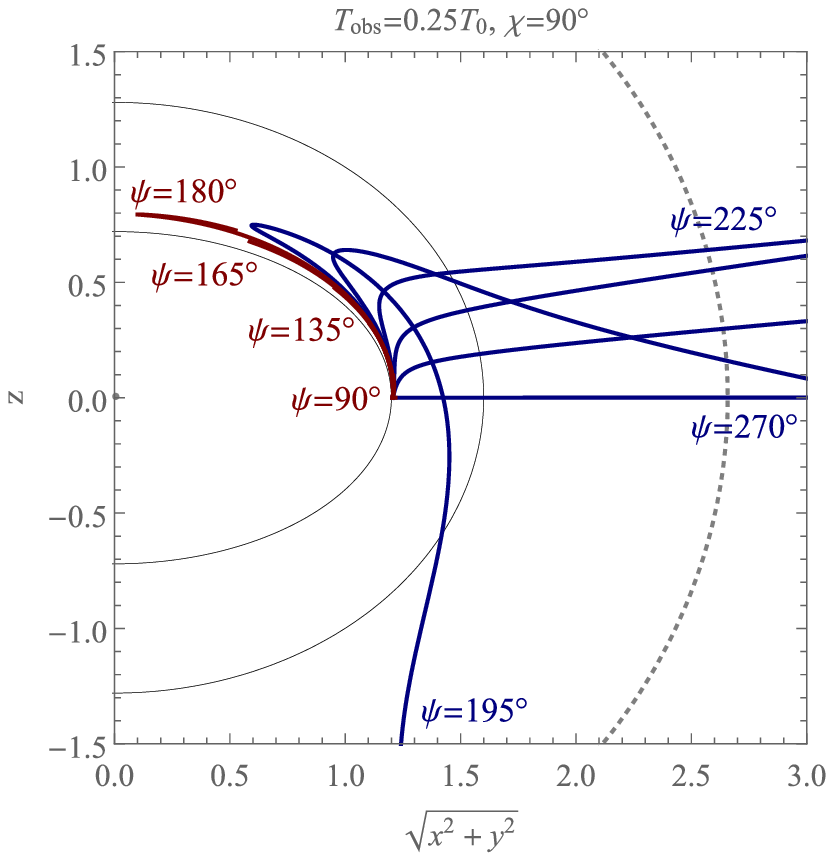}
    \end{minipage}\vspace{6.5pt}
    \begin{minipage}{0.45\textwidth}
        \includegraphics[width=0.95\textwidth]{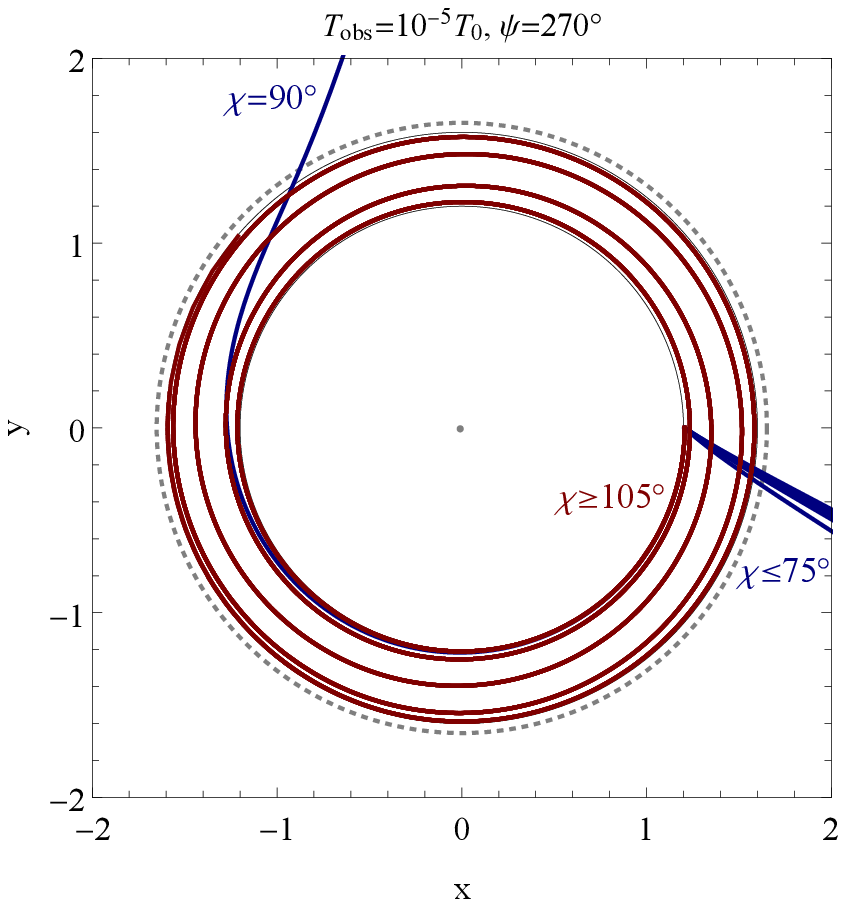}
    \end{minipage}
    \begin{minipage}{0.45\textwidth}
        \includegraphics[width=0.95\textwidth]{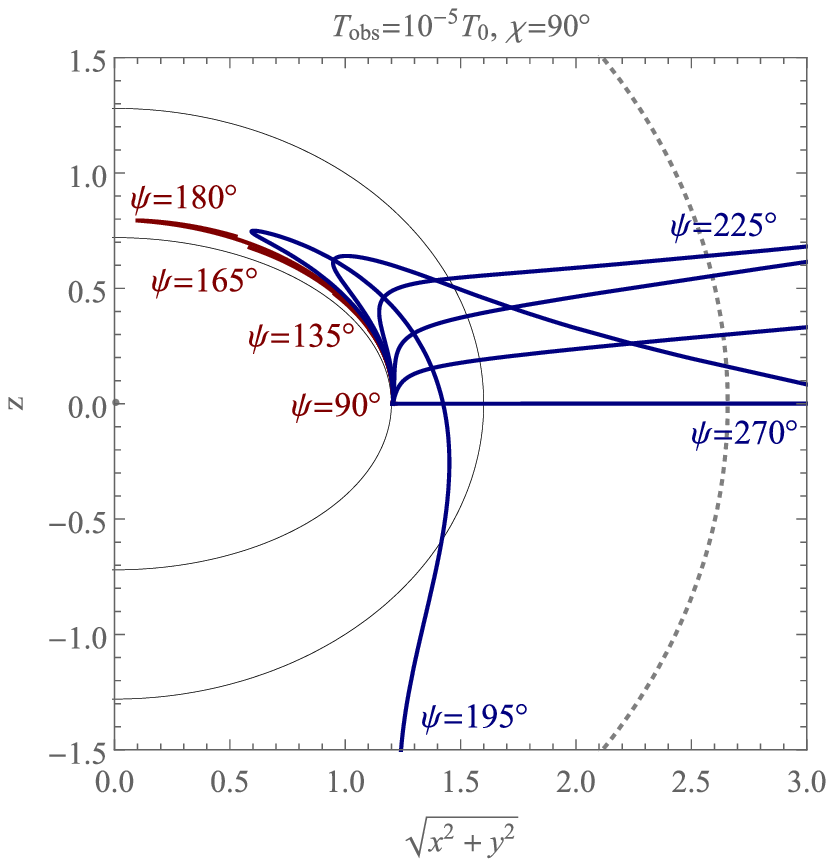}
    \end{minipage}%
    \caption{\label{fig:kerr_geodesics}(Color online). Null geodesics ray-traced backwards from an equatorial inflationary Kasner observer at ${T_\text{obs}=0.25T_0}$ (top two panels) and ${T_\text{obs}=10^{-5}T_0}$ (bottom two panels) to their Kerr origins. The left panels show a slice of the equatorial plane with Doran azimuthal coordinates, viewed from over the pole, and the right panels show a polar slice in co-rotating coordinates. In all panels, the two thin solid black curves shows the locations of the inner and outer horizons, and the dashed curves show the location of the null circular prograde equatorial (left) and polar (right) orbits. All geodesics are labelled by the viewing angle of the inflationary Kasner observer, equally spaced at intervals of $15^{\circ}$, and they originate either from the background at infinity (dark blue) or from the surface of the collapsing star (dark red). The parameters chosen here are ${u=0.02}$, ${r_1=0.73}$, ${\theta_1=90^{\circ}}$, and ${a=0.96}$.}
\end{figure*}

Some examples of photon paths reaching observers near the inner horizon are shown in Fig.~\ref{fig:kerr_geodesics}.  To avoid the effects of any coordinate singularities at the horizons, the paths are plotted using Doran coordinates, which are related to the Boyer-Lindquist coordinates by the transformations
\begin{subequations}
\begin{align}
    r_{\text{D}}&=r_{\text{BL}},\\
    dt_{\text{D}}&=dt_{\text{BL}}+\frac{\sqrt{2Mr(r^2+a^2)}}{r^2+a^2-2Mr}dr_{\text{BL}},\\
    \theta_{\text{D}}&=\theta_{\text{BL}},\\
    d\phi_{\text{D}}&=d\phi_{\text{BL}}+\frac{a\sqrt{2Mr/(r^2+a^2)}}{r^2+a^2-2Mr}dr_{\text{BL}}
\end{align}
\end{subequations}
\cite{dor00}. We limit our analysis to two equatorial observers, one in the inflation epoch (${T_\text{obs}=0.25T_0}$) and one deep into the collapse epoch (${T_\text{obs}=10^{-5}T_0}$); a more complete analysis of which photons arrive from which sources for different observer latitudes and radii is given in Sec.~\ref{sec:obs_experience}.

The two left panels of Fig.~\ref{fig:kerr_geodesics} show null geodesics in the equatorial plane, reaching an observer at ${(x,y)\approx(1.2,0)}$ just above the inner horizon. These geodesics are the continuation of the geodesics of Fig.~\ref{fig:ikasner_geodesics} when ${\psi=270^{\circ}}$---once the inflationary Kasner geodesics have been traced back from the observer to the point of matching at ${T=T_1}$, here they are continued backward in the Kerr metric to their point of origin at infinity (blue) or the outer horizon (red). As $\chi$ increases, the geodesics become more and more skewed until they asymptotically wrap around the photon sphere given by the dashed curve. The ${\chi=180^{\circ}}$ geodesic is omitted from the top left panel for simplicity; its form is identical to the ${\chi=180^{\circ}}$ geodesic in the lower left panel.

The behavior of the geodesics in the left two panels of Fig.~\ref{fig:kerr_geodesics} matches that of Fig.~\ref{fig:ikasner_geodesics}. In particular, when the observer has progressed deep into the collapse epoch (when ${T_{\text{obs}}\ll T_0}$), most light tends to focus along the principal null directions, so that most of the observer's field of view contains light originating from a small patch of the background (when ${\chi\le75^{\circ}}$) or illusory horizon (when ${\chi\ge105^{\circ}}$). In the collapse epoch, therefore, the observer sees most of the background sky squashed into a thin band close to ${\chi=90^{\circ}}$.

The right panels of Fig.~\ref{fig:kerr_geodesics} show geodesics for a fixed value of $\chi$ instead of $\psi$. Here, the observer is looking up and down instead of only looking within the equatorial plane. With this polar view, some geodesics (${\psi=195^{\circ}}$ to ${\psi=270^{\circ}}$) originate from infinity, but the others (${\psi=90^{\circ}}$ to ${\psi=180^{\circ}}$) originate at some arbitrary location below the outer horizon, where the collapsing star's surface existed at some point in the past. Though it may not be apparent from this view, these geodesics become increasingly skewed in the direction of the black hole's rotation as $\psi$ decreases, with the equatorial geodesic with ${\psi=90^{\circ}}$ occupying a single point in the polar view. Additionally, note that the geodesics in this right panel can be reflected across the ${z=0}$ line to obtain the geodesics for ${\psi<90^{\circ}}$ and ${\psi>270^{\circ}}$.

The polar null geodesics in the right two panels of Fig.~\ref{fig:kerr_geodesics} remain unchanged for an observer travelling from inflation to collapse, a consequence of the fact that the inflationary Kasner metric is isotropic in the $y$-$z$ plane, so that the dependence on $\psi$ in this case is trivial. Thus, an infalling equatorial Carter observer will see more and more of the sky flattening out and piling up toward the edges of the black hole's shadow, though the view at different altitudes will remain relatively unaffected by the inflationary Kasner metric.

\section{\label{sec:obs_experience}The Carter observer's experience}

As a brief caveat, it should be noted that the observer's field of view and the angles ${(\chi,\psi)}$ defined in this paper are completely dependent on the choice of tetrad frame. The interior Carter tetrad is adopted in this paper because of its simplicity and natural alignment with the principal null directions, but it is only a valid inertial rest frame for a free-falling observer below the outer horizon. In particular, an observer of mass $m$ at rest in the Carter frame must have orbital parameters ${E=0}$, ${L=0}$, and ${K=(ma\cos^2\!\theta)^2}$ (where $E$, $L$, and $K$ are defined analogously to Eq.~(\ref{eq:kerr_ELK}) but for a timelike geodesic). Nevertheless, a free-falling observer can decelerate to ${E=0}$ once they have passed through the outer horizon in order to stay at rest in the Carter frame and reproduce the results found here.

With that caveat out of the way, consider the complete field of view of a Carter observer during their descent into a black hole. The relevant object of analysis here is the black hole's shadow, the portion of the observer's sky void of any background photons. The perceived boundary between the black hole's shadow and the sky is determined by the location of the photon sphere, where photons circulate on a null, circular orbit for an indefinitely long amount of time before peeling off and reaching the observer. The orbital parameters of these photons (and the corresponding viewing angles) are given by the solutions to the equations
\begin{equation}
    R(r)=0,\qquad\frac{dR(r)}{dr}=0,
\end{equation}
parametrized by the allowed prograde ($-$) and retrograde (+) photon orbital radii, whose extremes are given by
\begin{equation}
    r_c=2M\left(1+\cos\!\left(\frac{2}{3}\cos^{-1}\!\left(\pm a\right)\right)\right)
\end{equation}
\cite{bar72}. Though we have been working with dimensionless Boyer-Lindquist coordinates, in this section we restore factors of $M$ to connect our equations to physical quantities.

\begin{figure*}
    \centering
    \begin{minipage}{0.95\textwidth}
        \includegraphics[width=0.8\textwidth]{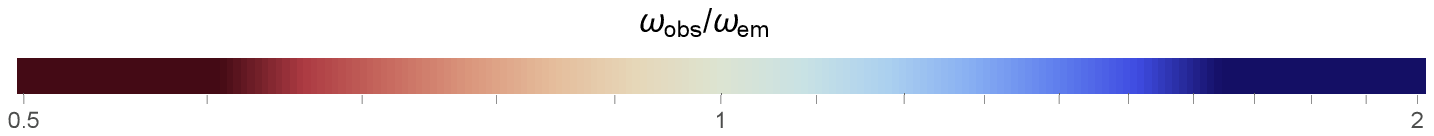}
    \end{minipage}\vspace{7pt}
    \begin{minipage}{0.45\textwidth}
        \includegraphics[width=0.95\textwidth]{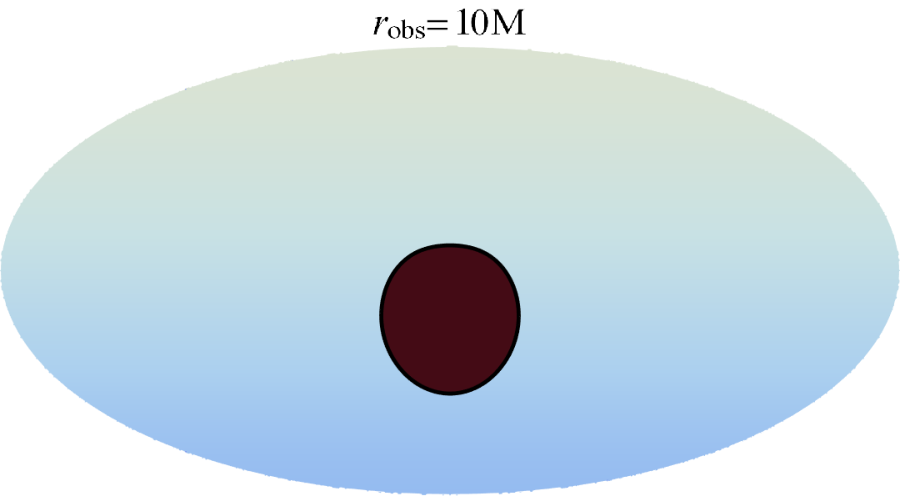}
    \end{minipage}
    \begin{minipage}{0.45\textwidth}
        \includegraphics[width=0.95\textwidth]{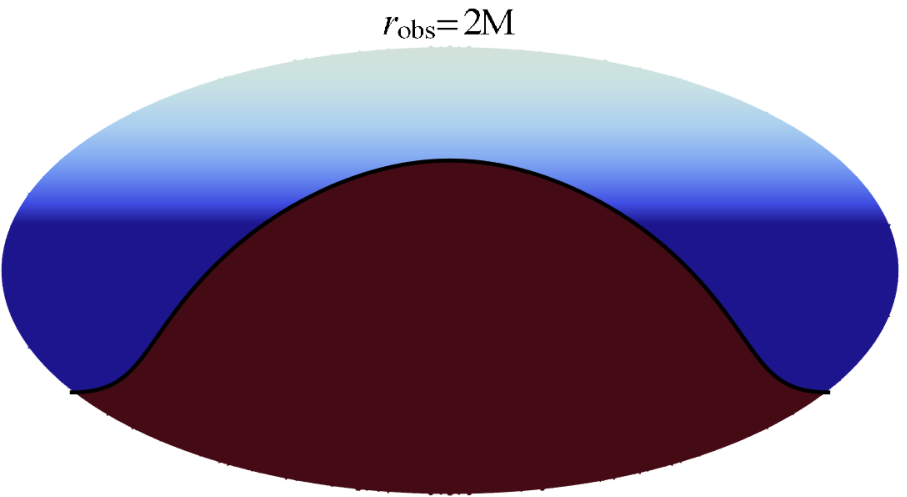}
    \end{minipage}\vspace{7pt}
    \begin{minipage}{0.45\textwidth}
        \includegraphics[width=0.95\textwidth]{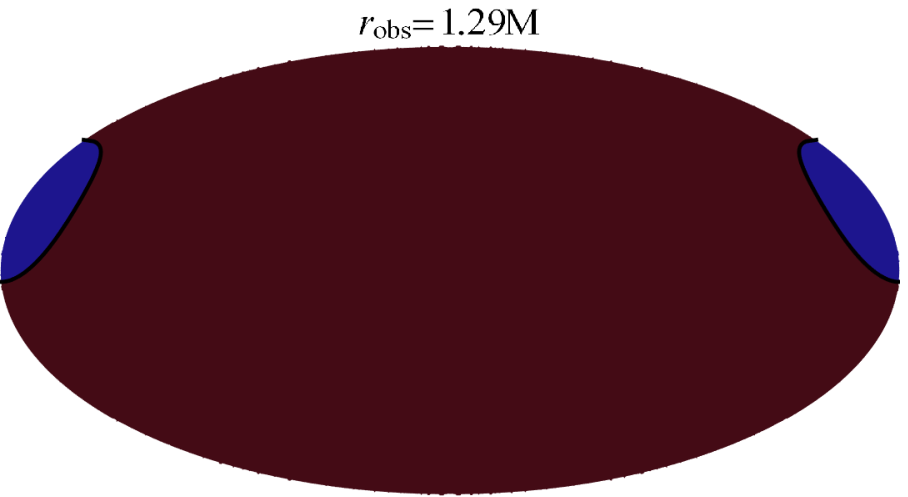}
    \end{minipage}
    \begin{minipage}{0.45\textwidth}
        \includegraphics[width=0.95\textwidth]{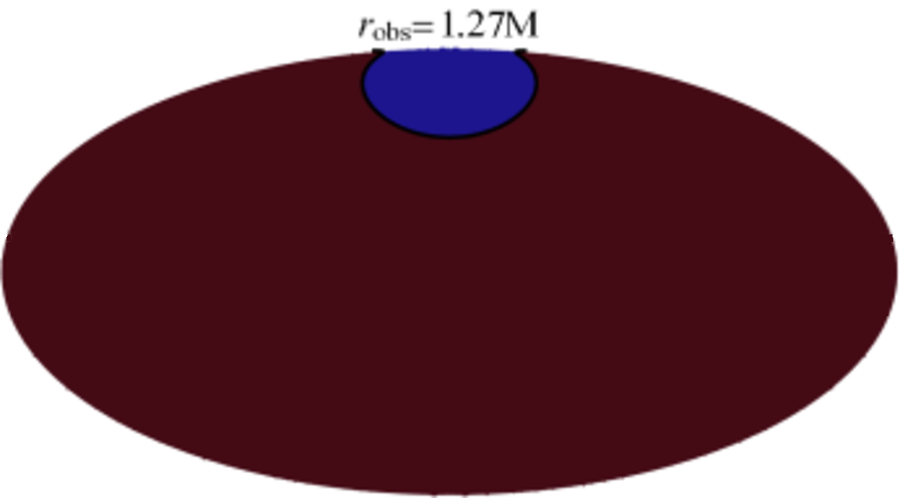}
    \end{minipage}\vspace{7pt}
    \begin{minipage}{0.45\textwidth}
        \includegraphics[width=0.95\textwidth]{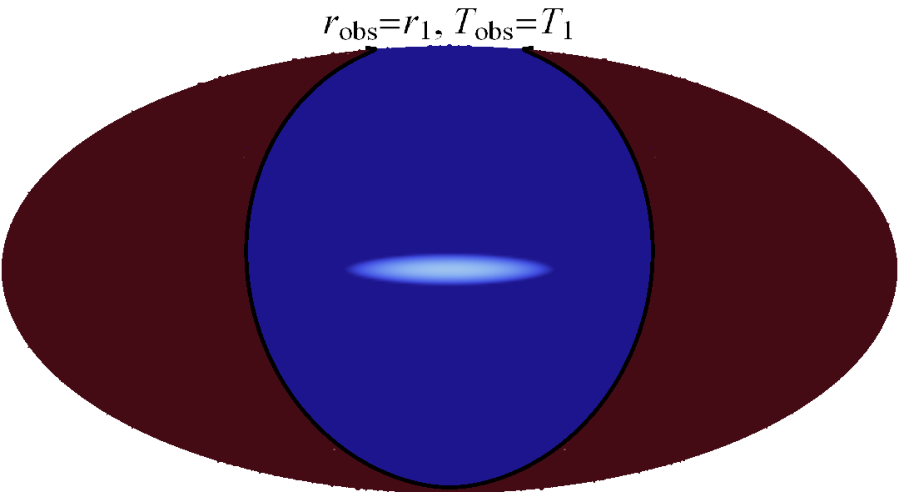}
    \end{minipage}
    \begin{minipage}{0.45\textwidth}
        \includegraphics[width=0.95\textwidth]{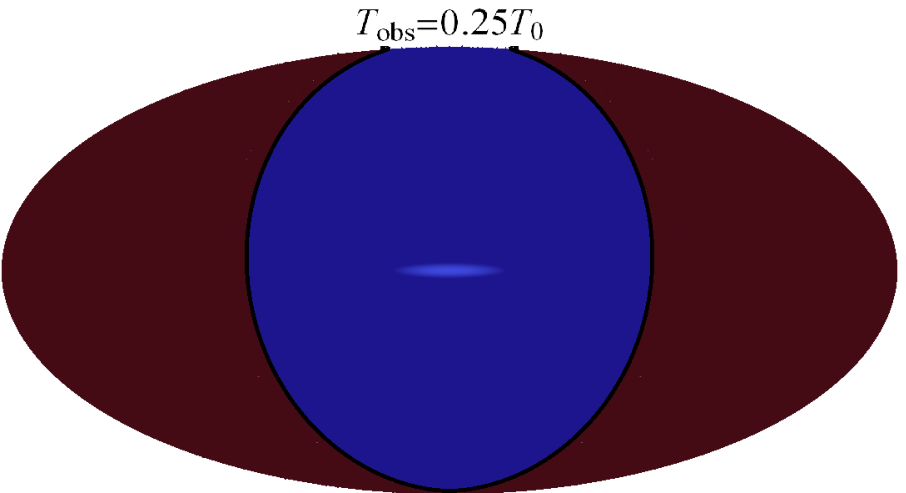}
    \end{minipage}\vspace{7pt}
    \begin{minipage}{0.45\textwidth}
        \includegraphics[width=0.95\textwidth]{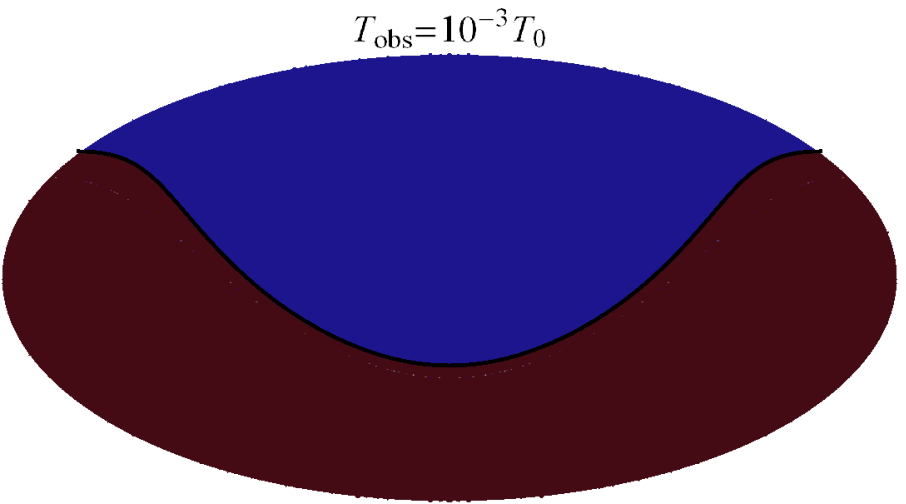}
    \end{minipage}
    \begin{minipage}{0.45\textwidth}
        \includegraphics[width=0.95\textwidth]{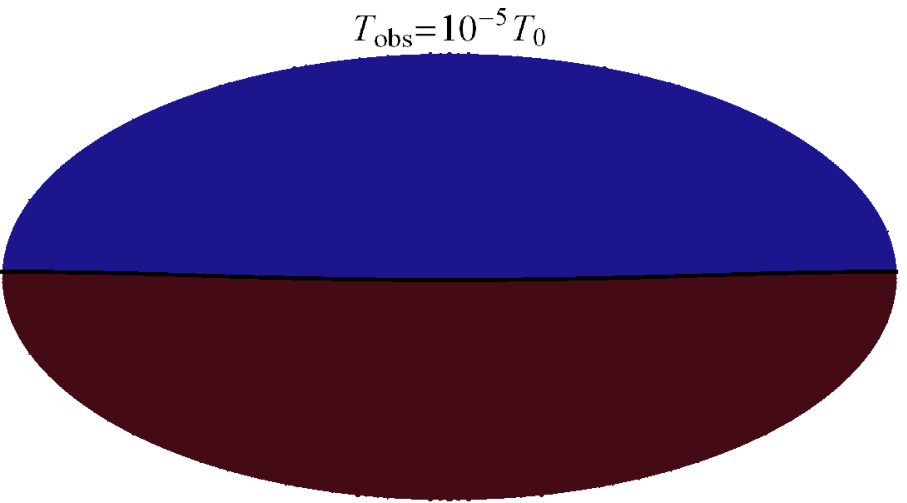}
    \end{minipage}%
    \caption{\label{fig:carter_shadow}(Color online). Mollweide projection of the full field of view of an infalling Carter observer in the equatorial plane at various radii and inflationary Kasner times recorded above each image. The black hole silhouette (black curve) separates the (generally) blueshifted photons sourced from ${r\to\infty}$ from the extremely redshifted photons sourced from ${r\approx r_+}$. The color represents the degree of redshift/blueshift. Note the change in the observer's orientation between the exterior (${r>1.28M}$) and interior (${r<1.28M}$) regions. The parameters used here are ${u=0.02}$, ${r_1=0.73}$, ${\theta_1=90^{\circ}}$, ${T_1\approx0.388T_0}$ and ${a=0.96}$.}
\end{figure*}

The black hole's shadow is shown in Fig.~\ref{fig:carter_shadow} for an equatorial observer at rest in the Carter frame at various radii and inflationary Kasner times. The observer's sky is displayed with a Mollweide projection, where the center corresponds to the observer's view directly ahead toward the black hole at ${\chi=180^{\circ}}$, the leftmost and rightmost points correspond to the view directly behind the observer at ${\chi=0^{\circ}}$, and the top and bottom points correspond to the view directly above (${\chi=90^{\circ},\ \psi=270^{\circ}}$) and below (${\chi=90^{\circ},\ \psi=90^{\circ}}$) the observer, respectively. More details about the projection are given in the Appendix. The black hole's spin axis is pointed to the right, so that the flow of spacetime is towards the observer above the shadow and away from the observer below the shadow.

The progression of images in Fig.~\ref{fig:carter_shadow} from top left to bottom right shows the view of the black hole as a Carter observer gets progressively closer to the inner horizon. Far from the black hole, the characteristic asymmetrical silhouette is seen in the top left image, with the background sky slightly blueshifted and the collapsing star's surface extremely redshifted (the color in these images is calculated from the energy component ${k^0}$ of the photon's four-momentum, normalized to ${k^0}$ at its point of origin). Then, as the observer approaches the outer horizon at ${r_+=1.28M}$ in the (non-inertial) exterior Carter frame, the shadow takes up more and more of the observer's view until the entire background sky is reduced to a single point behind the observer at the outer horizon. Then, as the Carter frame continues inward, the background sky behind the observer begins to grow again, until it takes up a little less than half the field of view once the observer reaches near the inner horizon (here ${r_-=0.72M}$).

As detailed in the Appendix, the field of view in Fig.~\ref{fig:carter_shadow} changes orientation between the exterior to the interior of the black hole. For ${r>1.28M}$, the black hole is in front of the observer and the sky is behind the observer, but for ${r<1.28M}$, we choose the black hole to be below the observer and the sky to be above, just as it is for the familiar case of an observer on the surface of Earth.

How does the inflationary Kasner solution modify the observer's view as they approach the inner horizon? The bottom two rows of Fig.~\ref{fig:carter_shadow} show a Carter observer's view in the inflation and collapse epochs. As inflation progresses, the black hole's shadow takes up approximately half of the equatorial observer's field of view, and the sky becomes more and more blueshifted. Then, as the observer continues into the collapse epoch, the black hole's shadow changes orientation until it appears as an infinite plane below the observer, taking up half of the field of view (in comparison, at a Schwarzschild singularity, an observer in free-fall also sees the shadow take up exactly half the field of view). Most of the sky above becomes squashed into a narrow band around ${\chi=90^{\circ}}$ (the horizontal midline in these images) as the observer approaches the inflationary Kasner singularity, as shown in the previous section. The validity of these images can be at least partially verified by comparing the points of origin of the geodesics of Fig.~\ref{fig:kerr_geodesics} in conjunction with their location in the images of Fig.~\ref{fig:carter_shadow} for ${T_{\text{obs}}=0.25T_0}$ and ${T_{\text{obs}}=10^{-5}T_0}$. To find the location of each ${(\chi,\psi)}$ point on the images of Fig.~\ref{fig:carter_shadow}, refer to Fig.~\ref{fig:chipsi_grids} in the Appendix.

\begin{figure}
    \centering
    \begin{minipage}{\columnwidth}
        \includegraphics[width=\textwidth]{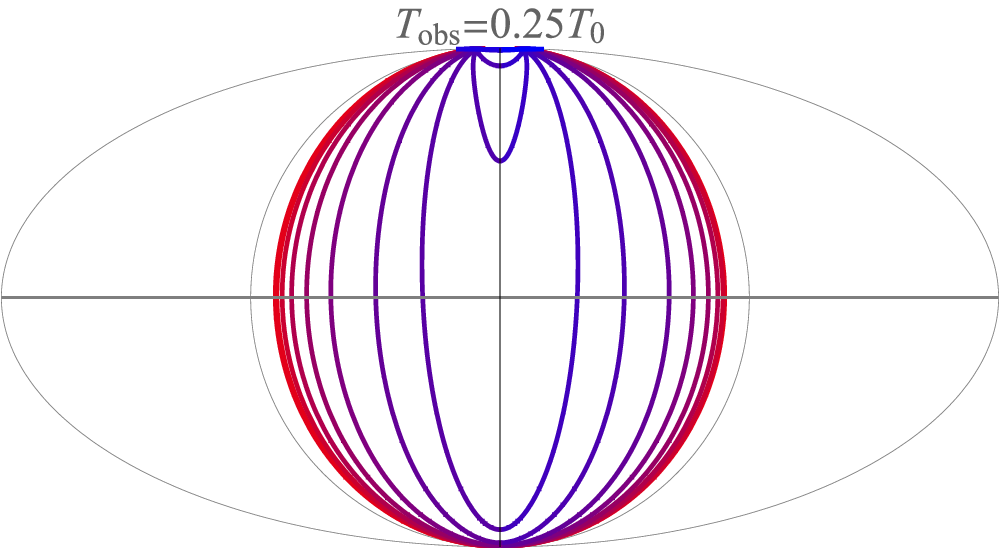}
    \end{minipage}\vspace{6.5pt}
    \begin{minipage}{\columnwidth}
        \includegraphics[width=\textwidth]{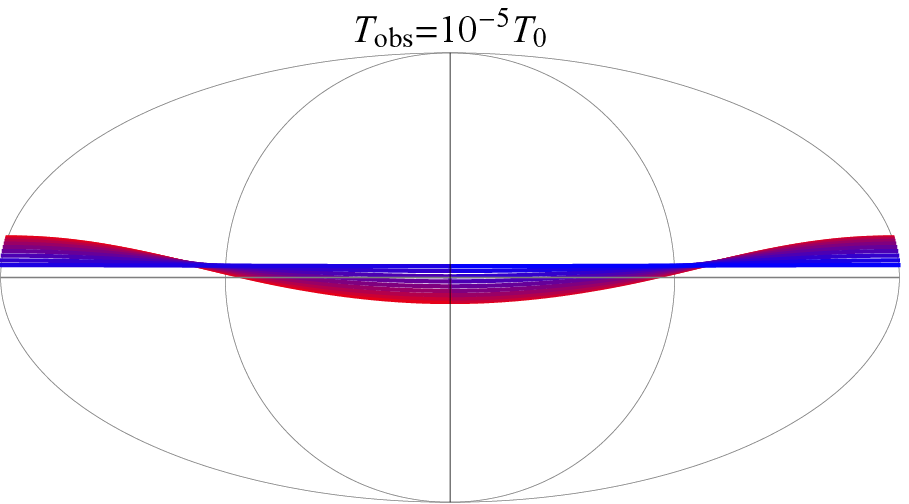}
    \end{minipage}%
    \caption{\label{fig:silhouettes} (Color online). Black hole silhouettes for an inflationary Kasner observer near the end of the inflation (upper panel) and collapse (lower panel) epochs, at a Boyer-Lindquist latitude ranging from the equator at ${\theta_\text{obs}=90^{\circ}}$ (blue) to the pole at ${\theta_\text{obs}=0^{\circ}}$ (red). The projection is the same as that of Figs.~\ref{fig:carter_shadow} and \ref{fig:chipsi_grids}b, and the parameters chosen here are ${u=0.02}$, ${r_1=0.73}$, and ${a=0.96}$.}
\end{figure}

What about observers outside of the equatorial plane? The final shape of the black hole's shadow depends on the Boyer-Lindquist latitude of the observer, as shown in Fig.~\ref{fig:silhouettes}. Near the end of the inflation epoch (top image), in the black hole's equatorial plane (${\theta_\text{obs}=90^{\circ}}$), the black hole takes up a little more than half of the observer's field of view. But at higher latitudes ($\theta_\text{obs}<90^{\circ}$), the shadow takes up more and more of the field of view, so that the sky in front of the observer appears as a thinner and thinner band connecting the principal null directions. Then, above some critical latitude, all photons must be ingoing, so that the shadow takes up the entire field of view at the end of the inflation epoch. An observer approaching the inner horizon at these latitudes close to the pole will see the sky constrict to a single point directly behind them. However, deep into the collapse epoch (bottom image), regardless of whether the observer is above or below the critical latitude, the black hole's shadow will always take up half the field of view below the observer, shifted $90^{\circ}$ from its location during the inflation epoch.

How much time passes for an observer experiencing the inflation and collapse of a black hole's inner horizon geometry? In the simplest case, for an equatorial observer of mass $m$ at rest in the interior Carter tetrad frame, the proper time that passes from the point of matching at ${T=T_1}$ to the singularity at ${T=0}$ is given by
\begin{equation}\label{eq:ikasner_tau}
    \tau=-M\int_{T_1}^0\frac{\text{e}^{(T-T_0)/2}}{\sqrt{16\pi\Phi_0T_0}}\left(\frac{T}{T_0}\right)^{-1/4}\!dT.
\end{equation}
For the parameters used in Fig.~\ref{fig:ikasner_evolution}, the proper time experienced by the observer is approximately
\begin{equation}
    \tau\approx\left(\frac{M}{M_{\astrosun}}\right)10^{-7}\text{ seconds},
\end{equation}
where ${M/M_{\astrosun}}$ is the mass of the black hole in units of solar masses. This proper time only changes by an order of magnitude or two at most across the physically valid domains of $a$, $\theta_0$, $r_0$, and $u$. In particular, the integral in Eq.~(\ref{eq:ikasner_tau}) approaches a constant value in the limit of an asymptotically small initial counter-streaming velocity $u$. However, in the same limit, the total time spent just in the collapse epoch (${T<1/2}$) becomes exponentially tiny (for the parameters used in Fig.~\ref{fig:ikasner_evolution} the time spent in the collapse epoch is already less than 1$\%$ of the time spent in the inflation epoch).

As a final note, the inflationary Kasner proper time calculated above is about an order of magnitude smaller than the proper time experienced by an equivalent observer in the Kerr spacetime travelling from the point of matching (${r=r_1}$) to the inner horizon (${r=r_-}$).

\section{\label{sec:conclusions}Conclusions}

The general classical outcome of the effect of accreted matter and radiation on a rotating black hole is the inflation and subsequent collapse of the spacetime near the inner horizon into a spacelike, BKL-like singularity. Here we have developed a simplified model that connects this collapsing geometry near the inner horizon to the Kerr exterior. The model, which we have called the inflationary Kasner model, is derived under the assumption that streams of matter near the inner horizon focus along the principal null directions at ultrarelativistic speeds, so that the Einstein tensor in the Carter frame approximately corresponds to that of a null, perfect fluid streaming at equal rates along the $x$-direction. Such an assumption leads to a Kasner-like form with two epochs, one corresponding to a purely radial collapse with Kasner exponents ${(1,0,0)}$, and a subsequent epoch with exponents ${(-\frac{1}{3},\frac{2}{3},\frac{2}{3})}$. The end result of the model is the termination of geodesics at a spacelike singularity at ${T=0}$; notably, the inner horizon and all the additional structure beyond it never get the chance to form.

We have verified the applicability of the inflationary Kasner metric to the near-inner horizon geometry of rotating, accreting black holes through comparison to a previously-derived solution, the conformally-separable Kerr metric. This solution comes equipped with a natural connection to the Kerr metric, along with a continuous evolution through the inflation and collapse epochs (and beyond, as has been shown computationally, through several BKL bounces \cite{ham17}). In the limit of asymptotically small accretion rates (${v\to0}$), the conformally-separable solution is equivalent to the inflationary Kasner solution during inflation and collapse, which lends credence to the validity of the latter model and allows for a more thorough interpretation of its parameters.

During the collapse epoch, the black hole's shadow spreads out to take up half of a Carter observer's field of view, exactly as in the Schwarzschild case, but the shadow is shifted ${90^{\circ}}$ from its position at the end of the classical Poisson-Israel mass inflation epoch, and unlike that latter case, the view is independent of the observer's latitude. Once the collapse epoch has proceeded long enough, the curvature will have diverged to such a large extent that the classical solution will surely break down. A calculation of the quantum back reaction will thus be necessary if one wishes to explore the spacetime evolution past this point in order to determine the final outcome of the collapse. The inflationary Kasner metric will hopefully provide a simpler basis for quantum calculations than more complicated models like the conformally-separable solution.

\appendix*
\section{}\label{sec:appendix}
\begin{figure}[b]
    \includegraphics[width=0.7\columnwidth]{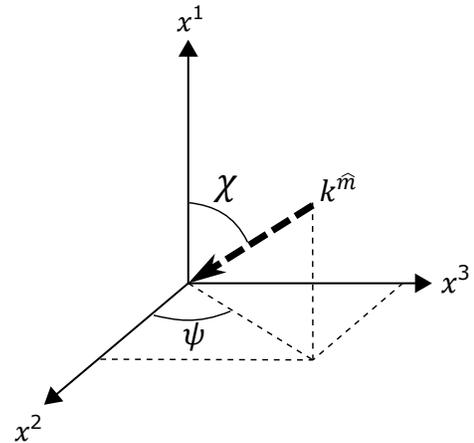}
    \caption{\label{fig:angle_defs} Definition of the viewing angles ${\chi}$ and ${\psi}$ with respect to the tetrad frame axes ${x^1}$, ${x^2}$, and ${x^3}$, for a photon with tetrad-frame four-momentum $k^{\hat{m}}$.}
\end{figure}

In this appendix we define and elaborate on the usage of the observer's viewing angles ${\chi}$ and ${\psi}$ employed throughout the paper. To determine the path of a null geodesic in a 3+1D spacetime uniquely, one needs to specify at most two constants of motion. For the inflationary Kasner metric, the spatial four-momentum components ${k_x}$, ${k_y}$, and ${k_z}$ uniquely label a null geodesic, but there is an extra degree of freedom associated with an arbitrary normalization factor for the four momentum's magnitude. Thus, we transform to a new set of constants that represents the celestial coordinates for an observer in the inflationary Kasner tetrad frame. In particular, the viewing angle ${\chi\in[0,\pi]}$ is defined to be the angle between the $x^1$-axis and ${-k_\text{IK}}$ (negative since the observer is seeing the photon reach the origin of their frame of reference), and ${\psi\in[0,2\pi)}$ is the angle between the $x^2$-axis and the projection of ${-k_\text{IK}}$ onto the $x^2$-$x^3$ plane. These viewing angles are shown in Fig.~\ref{fig:angle_defs}.

The definitions of the observer's viewing angles and their relation to the spatial covariant momenta via Eqs.~(\ref{eq:k_IK}) are given by
\begin{subequations}\label{eq:chipsiofki}
\begin{align}
    \tan\chi&\equiv\frac{\sqrt{\left(-k_{\text{IK}}^2\right)^2+\left(-k_{\text{IK}}^3\right)^2}}{-k_{\text{IK}}^1}\nonumber\\
    &=
    \frac{\sqrt{k_y^2+k_z^2}}{-k_x}\frac{a_{1}^{\text{obs}}}{a_{2}^{\text{obs}}},\\
    \tan\psi&\equiv\frac{-k_{\text{IK}}^3}{-k_{\text{IK}}^2}=\frac{-k_z}{-k_y},
\end{align}
\end{subequations}
or equivalently,
\begin{subequations}
\begin{align}
    k_x&=-\mathcal{E}a_1^{\text{obs}}\cos\chi,\\
    k_y&=-\mathcal{E}a_2^{\text{obs}}\sin\chi\cos\psi,\\
    k_z&=-\mathcal{E}a_2^{\text{obs}}\sin\chi\sin\psi,
\end{align}
\end{subequations}
where $a_1^{\text{obs}}$ and $a_2^{\text{obs}}$ are the values of the scales factors from Eq.~(\ref{eq:ikasner_scalefactors}) at time ${T=T_\text{obs}}$, and $\mathcal{E}$ is some positive normalization factor (the additional degree of freedom mentioned earlier).

\begin{figure}[ht]
    \centering
    \begin{minipage}{\columnwidth}
        \includegraphics[width=\textwidth]{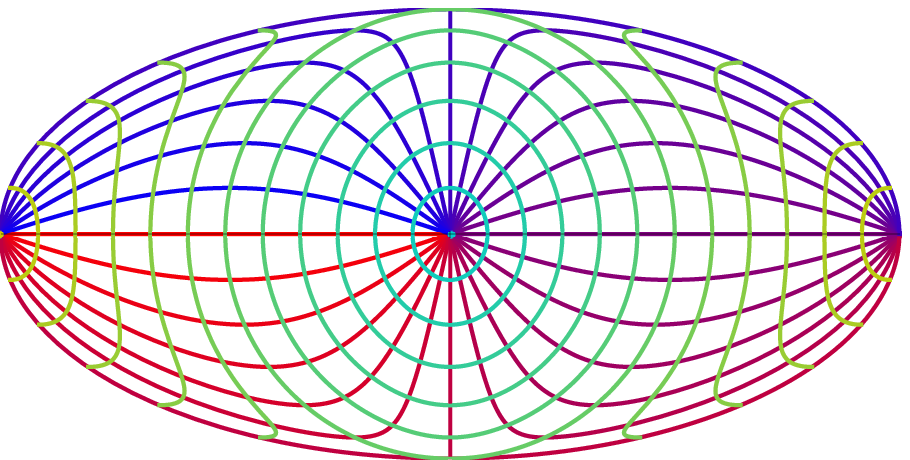}\\
        (a)
    \end{minipage}\vspace{6.5pt}
    \begin{minipage}{\columnwidth}
        \includegraphics[width=\textwidth]{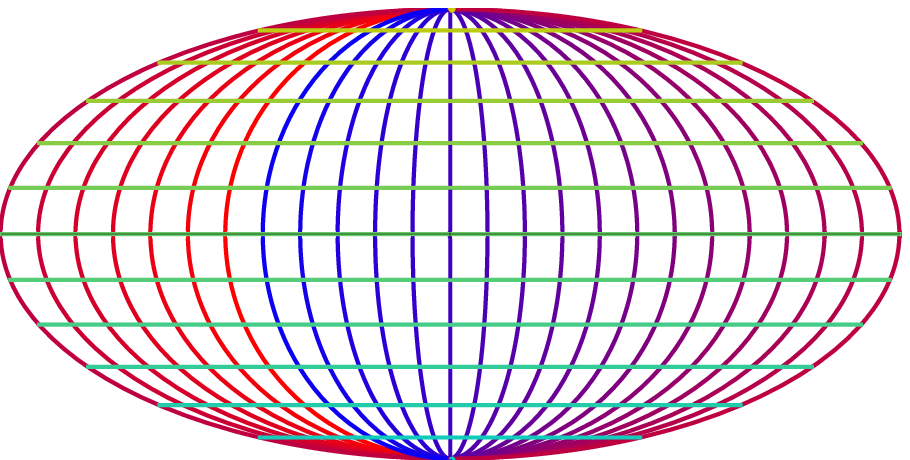}\\
        (b)
    \end{minipage}%
    \caption{\label{fig:chipsi_grids}(Color online). Coordinate grid of the viewing angles ${\chi}$ and ${\psi}$ on a Mollweide projection of the full field of view of an exterior (a) and interior (b) Carter observer. Lines of constant $\psi$ are equally spaced at 15$^{\circ}$ intervals from ${\psi=0^{\circ}}$ (red) to ${\psi=360^{\circ}}$ (blue), and lines of constant $\chi$ are equally spaced at 15$^{\circ}$ intervals from ${\chi=0^{\circ}}$ (yellow) to ${\chi=180^{\circ}}$ (cyan).}
\end{figure}

Physically, the $x^1$-axis of the tetrad frame is parallel to the principal null directions of the black hole, and the $x^2$-axis points in the ${\hat{\theta}}$ direction. When ${\chi=0^{\circ}}$, the observer is looking along the positive $x^1$-axis, away from the black hole, at ingoing photons. When ${\psi=0^{\circ}}$ and ${\chi=90^{\circ}}$, the observer is looking straight down along the positive $x^2$-axis, in the ${\hat{\theta}}$ direction. Geodesics with constant Boyer-Lindquist latitude are then given by ${\psi=90^{\circ}}$ and ${\psi=270^{\circ}}$.

When matching the tetrad-frame four-momenta of the inflationary Kasner solution with the Kerr solution at a Boyer-Lindquist radius of $r=r_0$, one can find the relation between the observer's viewing angles and the Kerr orbital parameters defined by Eqs.~(\ref{eq:kerr_ELK}). Inverting Eqs.~(\ref{eq:ELKofki}) and combining with Eqs.~(\ref{eq:chipsiofki}) yields
\begin{subequations}\label{eq:chipsiofELK}
\begin{align}
    \tan\chi&=\frac{\sqrt{K(-\Delta_{r_0})}}{E-\omega_{r_0}L}\frac{a_{1}^{\text{obs}}}{a_{2}^{\text{obs}}},\\
    \sin\psi&=\frac{\omega_{\theta_0}E-L}{\sqrt{K\Delta_{\theta_0}}}.
\end{align}
\end{subequations}

This relation holds for viewing angles defined for an observer in the interior Carter frame, within the event horizon. Outside the event horizon, the exterior Carter observer also possesses a set of viewing angles ${(\chi,\psi)}$, still defined by Fig.~\ref{fig:angle_defs}. However, those angles' relations to the Kerr orbital parameters will differ from the interior case, since the interior Carter frame differs from the exterior frame of Eqs.~(\ref{eq:cartertetrad}) in the swapping of ${e^0_{\ \mu}\leftrightarrow e^1_{\ \mu}}$ and ${\sqrt{-\Delta_r}\leftrightarrow\sqrt{+\Delta_r}}$. In the Kerr exterior, the viewing angles are related to the Kerr parameters by
\begin{subequations}\label{eq:chipsiofELK_ext}
\begin{align}
    \sin\chi&=\frac{\sqrt{K\Delta_{r_0}}}{E-\omega_{r_0}L},\\
    \sin\psi&=\frac{\omega_{\theta_0}E-L}{\sqrt{K\Delta_{\theta_0}}}.
\end{align}
\end{subequations}

The interior versus exterior region also differs in how we treat the Mollweide projections of the observer's sky, with reference to Figs.~\ref{fig:carter_shadow} and \ref{fig:silhouettes}. In the exterior region outside the event horizon, it is natural to choose the $-x^1$ direction (${\chi=180^{\circ}}$) for the center of the projection, since it corresponds to the direction toward the center of the black hole. However, in the interior region, it is more natural to choose the ${-x^3}$ direction (${\chi=90^{\circ}},\ \psi=270^{\circ}$) for the center of the projection, so that the black hole's shadow occupies the lower half of the field of view and the sky occupies the upper half, so as to coincide with our general notion of ``uprightness'' as we perceive of it on Earth.

In terms of the Mollweide projection's latitude ${\varphi\in[-\frac{\pi}{2},\frac{\pi}{2}]}$ and longitude ${\lambda\in[-\pi,\pi)}$ for Figs.~\ref{fig:carter_shadow} and \ref{fig:silhouettes}, defined by
\begin{subequations}
\begin{align}
    \varphi&=\sin^{-1}\!\left(\frac{2}{\pi}\left(y\sqrt{1-y^2}+\sin^{-1}\!y\right)\right),\\
    \lambda&=\frac{\pi}{2}\frac{x}{\sqrt{1-y^2}},
\end{align}
\end{subequations}
where ${x\in[-2,2]}$ and ${y\in[-1,1]}$, the viewing angles in the Kerr exterior are given by
\begin{subequations}
\begin{align}
    \tan\chi&=\frac{\sqrt{\sin^2\!\varphi+\cos^2\!\varphi\sin^2\!\lambda}}{-\cos\varphi\cos\lambda},\\
    \tan\psi&=\frac{\tan\varphi}{\sin\lambda},
\end{align}
\end{subequations}
and the viewing angles in the Kerr interior are given by
\begin{subequations}
\begin{align}
    \chi&=\frac{\pi}{2}-\varphi,\\
    \psi&=\left(\frac{3\pi}{2}-\lambda\right)\text{mod }2\pi.
\end{align}
\end{subequations}

For reference, Fig.~\ref{fig:chipsi_grids} shows the ${\chi-\psi}$ coordinate grid for both the interior and exterior Mollweide projection views of Figs.~\ref{fig:carter_shadow} and \ref{fig:silhouettes}.

\bibliography{apsbib}

\providecommand{\noopsort}[1]{}\providecommand{\singleletter}[1]{#1}%
\begin{thebibliography}{33}%
\makeatletter
\providecommand \@ifxundefined [1]{%
 \@ifx{#1\undefined}
}%
\providecommand \@ifnum [1]{%
 \ifnum #1\expandafter \@firstoftwo
 \else \expandafter \@secondoftwo
 \fi
}%
\providecommand \@ifx [1]{%
 \ifx #1\expandafter \@firstoftwo
 \else \expandafter \@secondoftwo
 \fi
}%
\providecommand \natexlab [1]{#1}%
\providecommand \enquote  [1]{``#1''}%
\providecommand \bibnamefont  [1]{#1}%
\providecommand \bibfnamefont [1]{#1}%
\providecommand \citenamefont [1]{#1}%
\providecommand \href@noop [0]{\@secondoftwo}%
\providecommand \href [0]{\begingroup \@sanitize@url \@href}%
\providecommand \@href[1]{\@@startlink{#1}\@@href}%
\providecommand \@@href[1]{\endgroup#1\@@endlink}%
\providecommand \@sanitize@url [0]{\catcode `\\12\catcode `\$12\catcode
  `\&12\catcode `\#12\catcode `\^12\catcode `\_12\catcode `\%12\relax}%
\providecommand \@@startlink[1]{}%
\providecommand \@@endlink[0]{}%
\providecommand \url  [0]{\begingroup\@sanitize@url \@url }%
\providecommand \@url [1]{\endgroup\@href {#1}{\urlprefix }}%
\providecommand \urlprefix  [0]{URL }%
\providecommand \Eprint [0]{\href }%
\providecommand \doibase [0]{https://doi.org/}%
\providecommand \selectlanguage [0]{\@gobble}%
\providecommand \bibinfo  [0]{\@secondoftwo}%
\providecommand \bibfield  [0]{\@secondoftwo}%
\providecommand \translation [1]{[#1]}%
\providecommand \BibitemOpen [0]{}%
\providecommand \bibitemStop [0]{}%
\providecommand \bibitemNoStop [0]{.\EOS\space}%
\providecommand \EOS [0]{\spacefactor3000\relax}%
\providecommand \BibitemShut  [1]{\csname bibitem#1\endcsname}%
\let\auto@bib@innerbib\@empty
\bibitem [{\citenamefont {Lifshitz}\ and\ \citenamefont
  {Khalatnikov}(1963)}]{lif63}%
  \BibitemOpen
  \bibfield  {author} {\bibinfo {author} {\bibfnamefont {E.}~\bibnamefont
  {Lifshitz}}\ and\ \bibinfo {author} {\bibfnamefont {I.}~\bibnamefont
  {Khalatnikov}},\ }\bibfield  {title} {\bibinfo {title} {Investigations in
  relativistic cosmology},\ }\href {https://doi.org/10.1080/00018736300101283}
  {\bibfield  {journal} {\bibinfo  {journal} {Advances in Physics}\ }\textbf
  {\bibinfo {volume} {12}},\ \bibinfo {pages} {185} (\bibinfo {year}
  {1963})}\BibitemShut {NoStop}%
\bibitem [{\citenamefont {Penrose}(1965)}]{pen65}%
  \BibitemOpen
  \bibfield  {author} {\bibinfo {author} {\bibfnamefont {R.}~\bibnamefont
  {Penrose}},\ }\bibfield  {title} {\bibinfo {title} {Gravitational collapse
  and space-time singularities},\ }\href
  {https://doi.org/10.1103/PhysRevLett.14.57} {\bibfield  {journal} {\bibinfo
  {journal} {Phys. Rev. Lett.}\ }\textbf {\bibinfo {volume} {14}},\ \bibinfo
  {pages} {57} (\bibinfo {year} {1965})}\BibitemShut {NoStop}%
\bibitem [{\citenamefont {Belinskii}\ \emph {et~al.}(1970)\citenamefont
  {Belinskii}, \citenamefont {Khalatnikov},\ and\ \citenamefont
  {Lifshitz}}]{bel70}%
  \BibitemOpen
  \bibfield  {author} {\bibinfo {author} {\bibfnamefont {V.}~\bibnamefont
  {Belinskii}}, \bibinfo {author} {\bibfnamefont {I.}~\bibnamefont
  {Khalatnikov}},\ and\ \bibinfo {author} {\bibfnamefont {E.}~\bibnamefont
  {Lifshitz}},\ }\bibfield  {title} {\bibinfo {title} {Oscillatory approach to
  a singular point in the relativistic cosmology},\ }\href
  {https://doi.org/10.1080/00018737000101171} {\bibfield  {journal} {\bibinfo
  {journal} {Advances in Physics}\ }\textbf {\bibinfo {volume} {19}},\ \bibinfo
  {pages} {525} (\bibinfo {year} {1970})}\BibitemShut {NoStop}%
\bibitem [{\citenamefont {Penrose}(1968)}]{pen68}%
  \BibitemOpen
  \bibfield  {author} {\bibinfo {author} {\bibfnamefont {R.}~\bibnamefont
  {Penrose}},\ }\bibfield  {title} {\bibinfo {title} {{Structure of
  space-time}},\ }in\ \href@noop {} {\emph {\bibinfo {booktitle} {{Battelle
  Rencontres}}}}\ (\bibinfo {year} {1968})\ pp.\ \bibinfo {pages}
  {121--235}\BibitemShut {NoStop}%
\bibitem [{\citenamefont {Simpson}\ and\ \citenamefont
  {Penrose}(1973)}]{sim73}%
  \BibitemOpen
  \bibfield  {author} {\bibinfo {author} {\bibfnamefont {M.}~\bibnamefont
  {Simpson}}\ and\ \bibinfo {author} {\bibfnamefont {R.}~\bibnamefont
  {Penrose}},\ }\bibfield  {title} {\bibinfo {title} {Internal instability in a
  {R}eissner-{N}ordstr{\"o}m black hole},\ }\href@noop {} {\bibfield  {journal}
  {\bibinfo  {journal} {Int.\ J.\ Theor.\ Phys.}\ }\textbf {\bibinfo {volume}
  {7}},\ \bibinfo {pages} {183} (\bibinfo {year} {1973})}\BibitemShut {NoStop}%
\bibitem [{\citenamefont {Poisson}\ and\ \citenamefont {Israel}(1990)}]{poi90}%
  \BibitemOpen
  \bibfield  {author} {\bibinfo {author} {\bibfnamefont {E.}~\bibnamefont
  {Poisson}}\ and\ \bibinfo {author} {\bibfnamefont {W.}~\bibnamefont
  {Israel}},\ }\bibfield  {title} {\bibinfo {title} {Internal structure of
  black holes},\ }\href {https://doi.org/10.1103/PhysRevD.41.1796} {\bibfield
  {journal} {\bibinfo  {journal} {Phys. Rev. D}\ }\textbf {\bibinfo {volume}
  {41}},\ \bibinfo {pages} {1796} (\bibinfo {year} {1990})}\BibitemShut
  {NoStop}%
\bibitem [{\citenamefont {Barrabes}\ \emph {et~al.}(1990)\citenamefont
  {Barrabes}, \citenamefont {Israel},\ and\ \citenamefont {Poisson}}]{bar90}%
  \BibitemOpen
  \bibfield  {author} {\bibinfo {author} {\bibfnamefont {C.}~\bibnamefont
  {Barrabes}}, \bibinfo {author} {\bibfnamefont {W.}~\bibnamefont {Israel}},\
  and\ \bibinfo {author} {\bibfnamefont {E.}~\bibnamefont {Poisson}},\
  }\bibfield  {title} {\bibinfo {title} {Collision of light-like shells and
  mass inflation in rotating black holes},\ }\href
  {https://doi.org/10.1088/0264-9381/7/12/002} {\bibfield  {journal} {\bibinfo
  {journal} {Classical and Quantum Gravity}\ }\textbf {\bibinfo {volume} {7}},\
  \bibinfo {pages} {L273} (\bibinfo {year} {1990})}\BibitemShut {NoStop}%
\bibitem [{\citenamefont {Ori}(1991)}]{ori91}%
  \BibitemOpen
  \bibfield  {author} {\bibinfo {author} {\bibfnamefont {A.}~\bibnamefont
  {Ori}},\ }\bibfield  {title} {\bibinfo {title} {Inner structure of a charged
  black hole: An exact mass-inflation solution},\ }\href
  {https://doi.org/10.1103/PhysRevLett.67.789} {\bibfield  {journal} {\bibinfo
  {journal} {Phys. Rev. Lett.}\ }\textbf {\bibinfo {volume} {67}},\ \bibinfo
  {pages} {789} (\bibinfo {year} {1991})}\BibitemShut {NoStop}%
\bibitem [{\citenamefont {Ori}(1992)}]{ori92}%
  \BibitemOpen
  \bibfield  {author} {\bibinfo {author} {\bibfnamefont {A.}~\bibnamefont
  {Ori}},\ }\bibfield  {title} {\bibinfo {title} {Structure of the singularity
  inside a realistic rotating black hole},\ }\href
  {https://doi.org/10.1103/PhysRevLett.68.2117} {\bibfield  {journal} {\bibinfo
   {journal} {Phys. Rev. Lett.}\ }\textbf {\bibinfo {volume} {68}},\ \bibinfo
  {pages} {2117} (\bibinfo {year} {1992})}\BibitemShut {NoStop}%
\bibitem [{\citenamefont {Dafermos}(2005)}]{daf05}%
  \BibitemOpen
  \bibfield  {author} {\bibinfo {author} {\bibfnamefont {M.}~\bibnamefont
  {Dafermos}},\ }\bibfield  {title} {\bibinfo {title} {The interior of charged
  black holes and the problem of uniqueness in general relativity},\ }\href
  {https://doi.org/10.1002/cpa.20071} {\bibfield  {journal} {\bibinfo
  {journal} {Communications on Pure and Applied Mathematics}\ }\textbf
  {\bibinfo {volume} {58}},\ \bibinfo {pages} {445} (\bibinfo {year}
  {2005})}\BibitemShut {NoStop}%
\bibitem [{\citenamefont {Brady}\ and\ \citenamefont {Chambers}(1995)}]{bra95}%
  \BibitemOpen
  \bibfield  {author} {\bibinfo {author} {\bibfnamefont {P.~R.}\ \bibnamefont
  {Brady}}\ and\ \bibinfo {author} {\bibfnamefont {C.~M.}\ \bibnamefont
  {Chambers}},\ }\bibfield  {title} {\bibinfo {title} {Nonlinear instability of
  {Kerr-type Cauchy} horizons},\ }\href
  {https://doi.org/10.1103/PhysRevD.51.4177} {\bibfield  {journal} {\bibinfo
  {journal} {Phys. Rev. D}\ }\textbf {\bibinfo {volume} {51}},\ \bibinfo
  {pages} {4177} (\bibinfo {year} {1995})}\BibitemShut {NoStop}%
\bibitem [{\citenamefont {Ori}\ and\ \citenamefont {Flanagan}(1996)}]{ori96}%
  \BibitemOpen
  \bibfield  {author} {\bibinfo {author} {\bibfnamefont {A.}~\bibnamefont
  {Ori}}\ and\ \bibinfo {author} {\bibfnamefont {E.~E.}\ \bibnamefont
  {Flanagan}},\ }\bibfield  {title} {\bibinfo {title} {How generic are null
  spacetime singularities?},\ }\href
  {https://doi.org/10.1103/PhysRevD.53.R1754} {\bibfield  {journal} {\bibinfo
  {journal} {Phys. Rev. D}\ }\textbf {\bibinfo {volume} {53}},\ \bibinfo
  {pages} {R1754} (\bibinfo {year} {1996})}\BibitemShut {NoStop}%
\bibitem [{\citenamefont {Ori}(1998)}]{ori98}%
  \BibitemOpen
  \bibfield  {author} {\bibinfo {author} {\bibfnamefont {A.}~\bibnamefont
  {Ori}},\ }\bibfield  {title} {\bibinfo {title} {Null weak singularities in
  plane-symmetric space-times},\ }\href
  {https://doi.org/10.1103/PhysRevD.57.4745} {\bibfield  {journal} {\bibinfo
  {journal} {Phys. Rev. D}\ }\textbf {\bibinfo {volume} {57}},\ \bibinfo
  {pages} {4745} (\bibinfo {year} {1998})}\BibitemShut {NoStop}%
\bibitem [{\citenamefont {Ori}(1999)}]{ori99}%
  \BibitemOpen
  \bibfield  {author} {\bibinfo {author} {\bibfnamefont {A.}~\bibnamefont
  {Ori}},\ }\bibfield  {title} {\bibinfo {title} {Oscillatory null singularity
  inside realistic spinning black holes},\ }\href
  {https://doi.org/10.1103/PhysRevLett.83.5423} {\bibfield  {journal} {\bibinfo
   {journal} {Phys. Rev. Lett.}\ }\textbf {\bibinfo {volume} {83}},\ \bibinfo
  {pages} {5423} (\bibinfo {year} {1999})}\BibitemShut {NoStop}%
\bibitem [{\citenamefont {Rubio}\ \emph {et~al.}(2021)\citenamefont {Rubio},
  \citenamefont {Filippo}, \citenamefont {Liberati}, \citenamefont {Pacilio},\
  and\ \citenamefont {Visser}}]{rub21}%
  \BibitemOpen
  \bibfield  {author} {\bibinfo {author} {\bibfnamefont {R.~C.}\ \bibnamefont
  {Rubio}}, \bibinfo {author} {\bibfnamefont {F.~D.}\ \bibnamefont {Filippo}},
  \bibinfo {author} {\bibfnamefont {S.}~\bibnamefont {Liberati}}, \bibinfo
  {author} {\bibfnamefont {C.}~\bibnamefont {Pacilio}},\ and\ \bibinfo {author}
  {\bibfnamefont {M.}~\bibnamefont {Visser}},\ }\href@noop {} {\bibinfo {title}
  {Inner horizon instability and the unstable cores of regular black holes}}
  (\bibinfo {year} {2021}),\ \Eprint {https://arxiv.org/abs/2101.05006}
  {arXiv:2101.05006 [gr-qc]} \BibitemShut {NoStop}%
\bibitem [{\citenamefont {Price}(1972)}]{pri72}%
  \BibitemOpen
  \bibfield  {author} {\bibinfo {author} {\bibfnamefont {R.~H.}\ \bibnamefont
  {Price}},\ }\bibfield  {title} {\bibinfo {title} {Nonspherical perturbations
  of relativistic gravitational collapse. {I. Scalar} and gravitational
  perturbations},\ }\href {https://doi.org/10.1103/PhysRevD.5.2419} {\bibfield
  {journal} {\bibinfo  {journal} {Phys. Rev. D}\ }\textbf {\bibinfo {volume}
  {5}},\ \bibinfo {pages} {2419} (\bibinfo {year} {1972})}\BibitemShut
  {NoStop}%
\bibitem [{\citenamefont {Hamilton}(2017)}]{ham17}%
  \BibitemOpen
  \bibfield  {author} {\bibinfo {author} {\bibfnamefont {A.~J.~S.}\
  \bibnamefont {Hamilton}},\ }\bibfield  {title} {\bibinfo {title} {Mass
  inflation followed by {Belinskii-Khalatnikov-Lifshitz} collapse inside
  accreting, rotating black holes},\ }\href
  {https://doi.org/10.1103/PhysRevD.96.084041} {\bibfield  {journal} {\bibinfo
  {journal} {Phys. Rev. D}\ }\textbf {\bibinfo {volume} {96}},\ \bibinfo
  {pages} {084041} (\bibinfo {year} {2017})}\BibitemShut {NoStop}%
\bibitem [{\citenamefont {Burko}(2002)}]{bur02}%
  \BibitemOpen
  \bibfield  {author} {\bibinfo {author} {\bibfnamefont {L.~M.}\ \bibnamefont
  {Burko}},\ }\bibfield  {title} {\bibinfo {title} {Survival of the black
  hole's {Cauchy} horizon under noncompact perturbations},\ }\href
  {https://doi.org/10.1103/PhysRevD.66.024046} {\bibfield  {journal} {\bibinfo
  {journal} {Phys. Rev. D}\ }\textbf {\bibinfo {volume} {66}},\ \bibinfo
  {pages} {024046} (\bibinfo {year} {2002})}\BibitemShut {NoStop}%
\bibitem [{\citenamefont {Burko}(2003)}]{bur03}%
  \BibitemOpen
  \bibfield  {author} {\bibinfo {author} {\bibfnamefont {L.~M.}\ \bibnamefont
  {Burko}},\ }\bibfield  {title} {\bibinfo {title} {Black-hole singularities: A
  new critical phenomenon},\ }\href
  {https://doi.org/10.1103/PhysRevLett.90.121101} {\bibfield  {journal}
  {\bibinfo  {journal} {Phys. Rev. Lett.}\ }\textbf {\bibinfo {volume} {90}},\
  \bibinfo {pages} {121101} (\bibinfo {year} {2003})}\BibitemShut {NoStop}%
\bibitem [{\citenamefont {Hamilton}\ and\ \citenamefont
  {Avelino}(2010)}]{ham10}%
  \BibitemOpen
  \bibfield  {author} {\bibinfo {author} {\bibfnamefont {A.~J.}\ \bibnamefont
  {Hamilton}}\ and\ \bibinfo {author} {\bibfnamefont {P.~P.}\ \bibnamefont
  {Avelino}},\ }\bibfield  {title} {\bibinfo {title} {The physics of the
  relativistic counter-streaming instability that drives mass inflation inside
  black holes},\ }\href {https://doi.org//10.1016/j.physrep.2010.06.002}
  {\bibfield  {journal} {\bibinfo  {journal} {Physics Reports}\ }\textbf
  {\bibinfo {volume} {495}},\ \bibinfo {pages} {1 } (\bibinfo {year}
  {2010})}\BibitemShut {NoStop}%
\bibitem [{\citenamefont {Hamilton}\ and\ \citenamefont
  {Polhemus}(2011)}]{ham11a}%
  \BibitemOpen
  \bibfield  {author} {\bibinfo {author} {\bibfnamefont {A.~J.~S.}\
  \bibnamefont {Hamilton}}\ and\ \bibinfo {author} {\bibfnamefont
  {G.}~\bibnamefont {Polhemus}},\ }\bibfield  {title} {\bibinfo {title}
  {Interior structure of rotating black holes. {I. Concise} derivation},\
  }\href {https://doi.org/10.1103/PhysRevD.84.124055} {\bibfield  {journal}
  {\bibinfo  {journal} {Phys. Rev. D}\ }\textbf {\bibinfo {volume} {84}},\
  \bibinfo {pages} {124055} (\bibinfo {year} {2011})}\BibitemShut {NoStop}%
\bibitem [{\citenamefont {Hamilton}(2011{\natexlab{a}})}]{ham11b}%
  \BibitemOpen
  \bibfield  {author} {\bibinfo {author} {\bibfnamefont {A.~J.~S.}\
  \bibnamefont {Hamilton}},\ }\bibfield  {title} {\bibinfo {title} {Interior
  structure of rotating black holes. {II. Uncharged} black holes},\ }\href
  {https://doi.org/10.1103/PhysRevD.84.124056} {\bibfield  {journal} {\bibinfo
  {journal} {Phys. Rev. D}\ }\textbf {\bibinfo {volume} {84}},\ \bibinfo
  {pages} {124056} (\bibinfo {year} {2011}{\natexlab{a}})}\BibitemShut
  {NoStop}%
\bibitem [{\citenamefont {Hamilton}(2011{\natexlab{b}})}]{ham11c}%
  \BibitemOpen
  \bibfield  {author} {\bibinfo {author} {\bibfnamefont {A.~J.~S.}\
  \bibnamefont {Hamilton}},\ }\bibfield  {title} {\bibinfo {title} {Interior
  structure of rotating black holes. {III. Charged} black holes},\ }\href
  {https://doi.org/10.1103/PhysRevD.84.124057} {\bibfield  {journal} {\bibinfo
  {journal} {Phys. Rev. D}\ }\textbf {\bibinfo {volume} {84}},\ \bibinfo
  {pages} {124057} (\bibinfo {year} {2011}{\natexlab{b}})}\BibitemShut
  {NoStop}%
\bibitem [{\citenamefont {Barcel\'o}\ \emph {et~al.}(2020)\citenamefont
  {Barcel\'o}, \citenamefont {Boyanov}, \citenamefont {Carballo-Rubio},\ and\
  \citenamefont {Garay}}]{bar20}%
  \BibitemOpen
  \bibfield  {author} {\bibinfo {author} {\bibfnamefont {C.}~\bibnamefont
  {Barcel\'o}}, \bibinfo {author} {\bibfnamefont {V.}~\bibnamefont {Boyanov}},
  \bibinfo {author} {\bibfnamefont {R.}~\bibnamefont {Carballo-Rubio}},\ and\
  \bibinfo {author} {\bibfnamefont {L.~J.}\ \bibnamefont {Garay}},\ }\bibfield
  {title} {\bibinfo {title} {{Black hole inner horizon evaporation in
  semiclassical gravity}},\ }\href@noop {} {\bibfield  {journal} {\bibinfo
  {journal} {{arXiv e-prints}}\ } (\bibinfo {year} {2020})},\ \Eprint
  {https://arxiv.org/abs/2011.07331} {arXiv:2011.07331 [gr-qc]} \BibitemShut
  {NoStop}%
\bibitem [{\citenamefont {{M\"uller}}\ and\ \citenamefont
  {{Grave}}(2009)}]{mul09}%
  \BibitemOpen
  \bibfield  {author} {\bibinfo {author} {\bibfnamefont {T.}~\bibnamefont
  {{M\"uller}}}\ and\ \bibinfo {author} {\bibfnamefont {F.}~\bibnamefont
  {{Grave}}},\ }\bibfield  {title} {\bibinfo {title} {Catalogue of
  spacetimes},\ }\href@noop {} {\bibfield  {journal} {\bibinfo  {journal}
  {arXiv e-prints}\ } (\bibinfo {year} {2009})},\ \Eprint
  {https://arxiv.org/abs/0904.4184} {arXiv:0904.4184 [gr-qc]} \BibitemShut
  {NoStop}%
\bibitem [{\citenamefont {Chandrasekhar}(1983)}]{cha83}%
  \BibitemOpen
  \bibfield  {author} {\bibinfo {author} {\bibfnamefont {S.}~\bibnamefont
  {Chandrasekhar}},\ }\href@noop {} {\emph {\bibinfo {title} {The mathematical
  theory of black holes}}}\ (\bibinfo  {publisher} {Clarendon Press},\ \bibinfo
  {address} {Oxford, England},\ \bibinfo {year} {1983})\BibitemShut {NoStop}%
\bibitem [{\citenamefont {Kasner}(1921)}]{kas21}%
  \BibitemOpen
  \bibfield  {author} {\bibinfo {author} {\bibfnamefont {E.}~\bibnamefont
  {Kasner}},\ }\bibfield  {title} {\bibinfo {title} {Geometrical theorems on
  {Einstein's} cosmological equations},\ }\href
  {http://www.jstor.org/stable/2370192} {\bibfield  {journal} {\bibinfo
  {journal} {American Journal of Mathematics}\ }\textbf {\bibinfo {volume}
  {43}},\ \bibinfo {pages} {217} (\bibinfo {year} {1921})}\BibitemShut
  {NoStop}%
\bibitem [{\citenamefont {Minguzzi}(2005)}]{min05}%
  \BibitemOpen
  \bibfield  {author} {\bibinfo {author} {\bibfnamefont {E.}~\bibnamefont
  {Minguzzi}},\ }\bibfield  {title} {\bibinfo {title} {The {Minkowski} metric
  in non-inertial observer radar coordinates},\ }\href
  {https://doi.org/10.1119/1.2060716} {\bibfield  {journal} {\bibinfo
  {journal} {American Journal of Physics}\ }\textbf {\bibinfo {volume} {73}},\
  \bibinfo {pages} {1117} (\bibinfo {year} {2005})}\BibitemShut {NoStop}%
\bibitem [{\citenamefont {Kasner}(1940)}]{kas40}%
  \BibitemOpen
  \bibfield  {author} {\bibinfo {author} {\bibfnamefont {E.}~\bibnamefont
  {Kasner}},\ }\href@noop {} {\emph {\bibinfo {title} {Mathematics and the
  imagination}}}\ (\bibinfo  {publisher} {Simon and Schuster},\ \bibinfo
  {address} {New York},\ \bibinfo {year} {1940})\BibitemShut {NoStop}%
\bibitem [{\citenamefont {Carter}(1968)}]{car68}%
  \BibitemOpen
  \bibfield  {author} {\bibinfo {author} {\bibfnamefont {B.}~\bibnamefont
  {Carter}},\ }\bibfield  {title} {\bibinfo {title} {Global structure of the
  {Kerr} family of gravitational fields},\ }\href
  {https://doi.org/10.1103/PhysRev.174.1559} {\bibfield  {journal} {\bibinfo
  {journal} {Phys. Rev.}\ }\textbf {\bibinfo {volume} {174}},\ \bibinfo {pages}
  {1559} (\bibinfo {year} {1968})}\BibitemShut {NoStop}%
\bibitem [{\citenamefont {{Thorne}}(1974)}]{tho74}%
  \BibitemOpen
  \bibfield  {author} {\bibinfo {author} {\bibfnamefont {K.~S.}\ \bibnamefont
  {{Thorne}}},\ }\bibfield  {title} {\bibinfo {title} {Disk-accretion onto a
  black hole. {II. Evolution} of the hole},\ }\href
  {https://doi.org/10.1086/152991} {\bibfield  {journal} {\bibinfo  {journal}
  {\apj}\ }\textbf {\bibinfo {volume} {191}},\ \bibinfo {pages} {507} (\bibinfo
  {year} {1974})}\BibitemShut {NoStop}%
\bibitem [{\citenamefont {Doran}(2000)}]{dor00}%
  \BibitemOpen
  \bibfield  {author} {\bibinfo {author} {\bibfnamefont {C.}~\bibnamefont
  {Doran}},\ }\bibfield  {title} {\bibinfo {title} {New form of the {Kerr}
  solution},\ }\href {https://doi.org/10.1103/PhysRevD.61.067503} {\bibfield
  {journal} {\bibinfo  {journal} {Phys. Rev. D}\ }\textbf {\bibinfo {volume}
  {61}},\ \bibinfo {pages} {067503} (\bibinfo {year} {2000})}\BibitemShut
  {NoStop}%
\bibitem [{\citenamefont {{Bardeen}}\ \emph {et~al.}(1972)\citenamefont
  {{Bardeen}}, \citenamefont {{Press}},\ and\ \citenamefont
  {{Teukolsky}}}]{bar72}%
  \BibitemOpen
  \bibfield  {author} {\bibinfo {author} {\bibfnamefont {J.~M.}\ \bibnamefont
  {{Bardeen}}}, \bibinfo {author} {\bibfnamefont {W.~H.}\ \bibnamefont
  {{Press}}},\ and\ \bibinfo {author} {\bibfnamefont {S.~A.}\ \bibnamefont
  {{Teukolsky}}},\ }\bibfield  {title} {\bibinfo {title} {Rotating black holes:
  Locally nonrotating frames, energy extraction, and scalar synchrotron
  radiation},\ }\href {https://doi.org/10.1086/151796} {\bibfield  {journal}
  {\bibinfo  {journal} {\apj}\ }\textbf {\bibinfo {volume} {178}},\ \bibinfo
  {pages} {347} (\bibinfo {year} {1972})}\BibitemShut {NoStop}%
\end{thebibliography}%

\end{document}